\documentclass[final,3p,times]{elsarticle}

\usepackage{lineno,hyperref}
\modulolinenumbers[10]
\usepackage[fleqn]{amsmath}
\usepackage{bm}
\usepackage{physics}
\usepackage{subcaption}
\usepackage{color}
\usepackage{import}
\graphicspath{{figures/}}
\usepackage{booktabs} 
\journal{}

\DeclareMathOperator*{\minimize}{minimize}

\biboptions{authoryear}

\begin{document}

\begin{frontmatter}

\title{Topology optimization of force densities for form finding of cable structures}


\author[mymainaddress]{Nicol\`{o} Pollini}
\ead{nicolo@technion.ac.il}

\address[mymainaddress]{Faculty of Civil and Environmental Engineering, Technion - Israel Institute of Technology, Haifa, Israel}

\begin{abstract}
This paper presents a topology optimization approach for the form finding of cable networks based on the force density method. The proposed framework combines the force density method with topology optimization based on continuous density variables to identify the equilibrated geometry and effective connectivity of cable structures simultaneously. Design variables are assigned to the cable network members and control their force densities through a SIMP interpolation, complemented by an explicit binary promoting term in the objective function. As a result, cable members vanish or remain active throughout the optimization process. The equilibrium configuration is obtained by enforcing the force density method equations, while the objective drives the optimized network as close as possible to a prescribed reference form. Volume constraints, passive boundary regions, and a minimum connectivity requirement at loaded joints are incorporated to yield physically meaningful and numerically stable designs. Several numerical examples based on cable net ground structures defined by horizontal, vertical, and diagonal members are used to study the proposed approach. A dedicated rounding study shows that the relaxed density variables converge to nearly discrete values at the end of the optimization process, with only a minor correction required after thresholding. A study of the volume fraction constraint further shows that reducing the permitted structural volume can trigger a qualitative transition in the optimized topology, from a redundant, densely braced configuration to a minimal load path. Finally, an assessment of solver performance shows that the proposed formulation scales reasonably well with problem size. The results demonstrate that the proposed approach generates innovative and interpretable cable network forms using a limited amount of structural material and provides a basis for future extensions.
\medskip

\noindent\textbf{Graphic abstract}

\bigskip

\noindent\centering\includegraphics[width=0.8\textwidth]{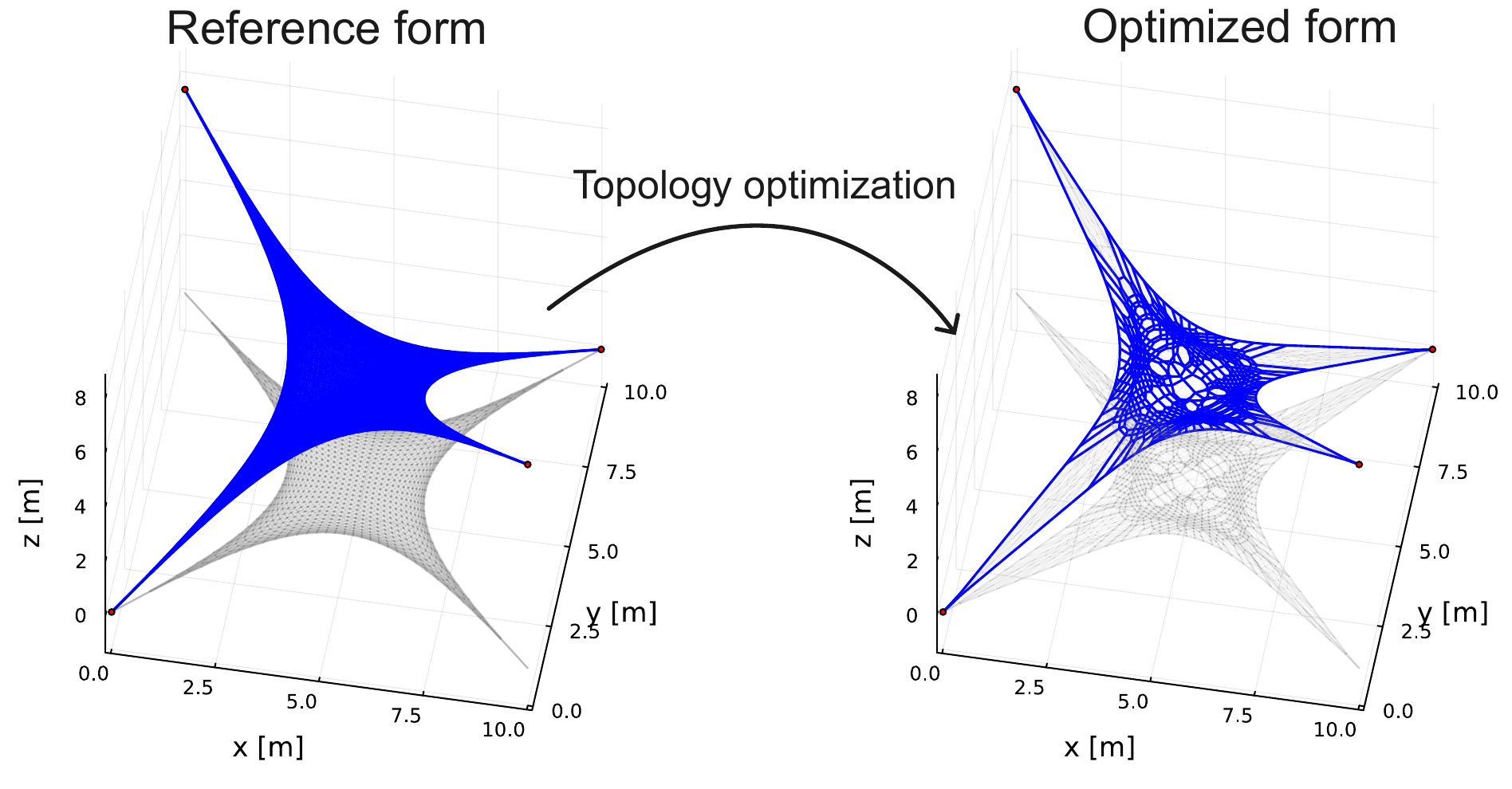}
\end{abstract}


\end{frontmatter}



\section{Introduction}
\label{sec:intro}

Lightweight cable structures are widely studied, both in research and in practice, for their strong aesthetic appeal, their ability to cover large spans and areas with minimal material, and their overall structural efficiency. These structural systems are composed of cable elements that resist only axial tensile forces. Consequently, the geometric layout, or form, of such structures is intimately connected to the internal forces carried by their elements, which typically arise from a combination of prestressing and external loads. This represents a fundamental departure from conventional civil and structural engineering systems, in which the geometry is prescribed a priori and the internal forces are subsequently derived from the fixed geometry under the applied loads and other actions. In cable structures, by contrast, geometry and internal forces are tightly coupled and must be determined simultaneously through what is commonly referred to as form finding \citep{adriaenssens2014shell,chiang2022form,miki2025almost}. A classical approach to the form finding of lightweight cable structures is the Force Density Method (FDM), originally proposed by Schek in 1974 \citep{Schek1974}. In this method, the boundary conditions are imposed by prescribing the locations of the fixed nodes, and a force density is assigned to each element or group of elements. By writing the equilibrium equations at each free node and introducing the force densities as prescribed parameters, the nonlinear form finding problem reduces to a linear system of equations that can be solved directly for the $x$, $y$, and $z$ coordinates of the free nodes. The literature also contains an earlier reference by \citet{linkwitz1971einige}, often cited as the origin of the FDM, although it is written in German.
In this context, Argyris et al.\ developed a finite element approach to form finding \citep{argyris1974general,tabarrok1992nonlinear}. Shortly thereafter, in 1975, Barnes proposed the dynamic relaxation method \citep{barnes1975application,barnes1988form,barnes1999form,adriaenssens2012finding}, which determines the equilibrium configuration of a tensile structure through a dynamic analysis performed incrementally in time with kinematic damping.
For a comparative review of form finding methods, the interested reader is referred to \citet{veenendaal2012overview}, and references therein. Veenendaal and Block present a unified computational framework for form finding, in which well known methods such as the FDM, dynamic relaxation, and the updated reference strategy are recast using a common notation and compared.

The FDM is one of the most popular and widely used methods for the form finding of cable structures. In his original work, Schek also discussed the possibility of extending the method to a nonlinear version that allows additional requirements to be imposed, such as prescribed relative distances between selected nodes, prescribed tensile forces in the elements, and possibly prescribed initial undeformed lengths \citep{Schek1974}. In 2012, \citet{malerba2012extended} proposed the Extended FDM, which made it possible to prescribe the positions and magnitudes of reactions at selected fixed nodes. The Extended FDM also accommodates mixed structures composed of tensioned cables and compressed struts.
Building on this extension, \citet{quagliaroli2013flexible} further specialized the Extended FDM to cable nets interacting with members exhibiting flexural behavior. Given a cable assembly and a prescribed loading condition, the aim is to determine the pretensioning system that replaces both the static and kinematic functions of the internal reactions of a flexural elastic continuous beam, as is the case, for instance, of bridge decks suspended by cables. \citet{pauletti2008natural} introduced the Natural FDM for the shape finding of cable and membrane structures. Their formulation preserves the linearity of the original FDM while accommodating irregular triangular membrane meshes and, when applied iteratively, can generate minimal surfaces associated with uniform isotropic plane stress states. More recently, \citet{pauletti2024extension} extended the Natural FDM to three-dimensional problems.
\citet{auricchio2026extended} extended the FDM to the form finding of gridshells with bending resistance in the vertical plane. Their formulation expands the force density concept beyond axial forces to account for shear forces and bending moments, allowing the equilibrium form to be determined for gridshell members that do not behave as purely axial elements. More recently, \citet{bruggi2026form} incorporated kinematic relations and a linear elastic constitutive law into an FDM framework for compressive gridshells. The resulting constrained optimization formulation controls both the structural form and elastic response while minimizing reaction forces or compliance.
The FDM has also been extended to the form finding of compression only structures, including concrete and masonry systems \citep{panozzo2013designing,marmo2017reformulation,avelino2021assessing}.
These developments further illustrate the versatility of force density approaches in supporting the design of new, unconventional structural forms.

Several studies have explored the integration of optimization methods into the form finding process \citep{hayashi2019fdmopt,zhang2021formfinding}. Some approaches use gradient free algorithms, including genetic algorithms \citep{pugnale2007morphogenesis,koohestani2012form}.
Beyond cable nets, form finding and optimization have also been combined in the design of gridshell structures \citep{qin2020genetic}. For example, \citet{melchiorre2025form} coupled the improved Multibody Rope Approach with multiobjective metaheuristic optimization (NSGA-II) to minimize material use, construction complexity, and material waste simultaneously during the preliminary design of gridshells. This work further illustrates the growing interest in coupling form finding with structural and manufacturing objectives in the optimization of lightweight structural systems.
Although effective, these methods can be computationally expensive because population based algorithms (such as genetic algorithms) generally require a large number of function evaluations \citep{martins2021engineering}. Other approaches use gradient based algorithms that exploit first order derivative information to solve the optimization problem \citep{zhang2006adaptive}. For example, \citet{ohsaki2017force} proposed a method for simultaneously optimizing the geometry and topology of plane and spatial trusses, using force densities as design variables to minimize compliance under a prescribed volume constraint. 
\citet{jiang2018form} proposed two form finding methods for gridshell design: a potential energy method, which determines the equilibrium form by minimizing the total potential energy of the system, and an augmented ground structure method, in which the load application points are treated as additional design variables. The two methods and the classical FDM were compared through several numerical examples. 
Similarly, \citet{zhong2025design} proposed a gradient based shape optimization framework for gridshells that uses nodal coordinates as design variables and first order sensitivity information to reduce structural compliance. Shape optimization is combined with a remeshing procedure to improve triangular panel similarity and manufacturability while retaining structural performance.
More recently, \citet{pastrana2026differentiable} proposed the differentiable FDM (DFDM), which uses automatic and analytical differentiation to enable gradient based optimization of funicular geometries subject to architectural, manufacturing, and structural constraints. The DFDM extends the classical FDM to pin jointed bar systems subject to shape dependent loads by embedding an iterative equilibrium solver within a fully differentiable framework. The DFDM thus exemplifies a broader trend toward embedding form finding based on force densities within gradient based optimization frameworks. This trend is closely related to the present work, although the objective differs. While the DFDM performs shape optimization for a fixed connectivity, the present work concurrently optimizes the equilibrium geometry and connectivity of the cable network.
\citet{liew2018load} and subsequently \citet{liew2020constrained} proposed a constrained FDM for the optimization of compression only shells and gridshells, treating the force densities as continuous design variables. The gradient based formulation employs analytical derivatives to minimize structural material volume and control member lengths while imposing constraints on forces, force densities, and nodal movements. Liew's method optimizes the force density distribution of a network with prescribed and fixed connectivity. Although some elements may acquire small force densities, the existence of an element is not treated as an explicit discrete design decision.
More recently, \citet{he2025minimum} extended numerical layout optimization to determine the form and force flow topology of compression only vaults. The underlying convex formulation was originally introduced by \citet{bolbotowski2022optimal}, although its presentation, as stated by He et al., relied heavily on mathematical proofs and perhaps an auxiliary abstract formulation. He et al.\ reinterpreted the method through accessible engineering concepts, derived the conic programming problem directly from classical truss layout optimization, and demonstrated a workflow that progresses from benchmark solutions based on a complete ground structure to more practical designs obtained by restricting member connectivity.

This paper presents a novel two stage approach to form finding based on a topology optimization framework for cable networks \citep{bendsoe2004topology,sigmund2013topology}. In the first stage, a reference equilibrium form is obtained using either the linear or nonlinear FDM. As an additional contribution, the nonlinear FDM originally introduced by Schek is recast as a constrained optimization problem that can be formulated and solved efficiently using a modern programming language and standard computational resources. In the second stage, the reference cable network is topology optimized by assigning a binary, pseudo-density, design variable to each member to indicate whether it is removed or retained. The resulting discrete problem is relaxed into a continuous formulation, in which a SIMP interpolation \citep{bendsoe1999material} penalizes intermediate pseudo-density values. An explicit binary promoting term in the objective function further drives the solution toward a crisp, nearly discrete design whose variables approach zero or one. The resulting topology optimization problem is solved with a nonlinear programming algorithm that identifies new equilibrium forms that remain as close as possible to the reference configuration. Volume constraints, passive boundary regions, and a minimum connectivity requirement at loaded joints are incorporated to yield physically meaningful and numerically stable structural forms. When the reference form is generated using the nonlinear FDM, additional geometric constraints may also be retained during topology optimization, including requirements on relative nodal positions, distances between nodes, and prescribed member lengths. Numerical studies assess the influence of the prescribed volume fraction on the resulting structural form and topology, and investigate the effect of problem size on computational cost.

The remainder of this article is organized as follows. Section~\ref{sec:fdm} reviews the linear and nonlinear FDM formulations. Section~\ref{sec:topopt} presents the integer topology optimization problem, its continuous relaxation, and the computational implementation. Section~\ref{sec:numex} illustrates and assesses the proposed approach through numerical examples, including studies of rounding, the volume fraction constraint, computational performance, and topology optimization with additional geometric constraints inherited from the nonlinear FDM. Finally, Section~\ref{sec:end} summarizes the main conclusions.


\section{Overview of the FDM}
\label{sec:fdm}
The FDM is described in detail by \citet{Schek1974}. A brief description is provided here because its mathematical formulation is tightly coupled to the topology optimization approach presented in this paper and clarifies the overall methodology.

\subsection{Linear FDM}
\label{subsec:linfdm}

\begin{figure*}[htbp!]
    \centering

    \begin{subfigure}{0.475\textwidth}
        \centering
        \includegraphics[width=\textwidth]{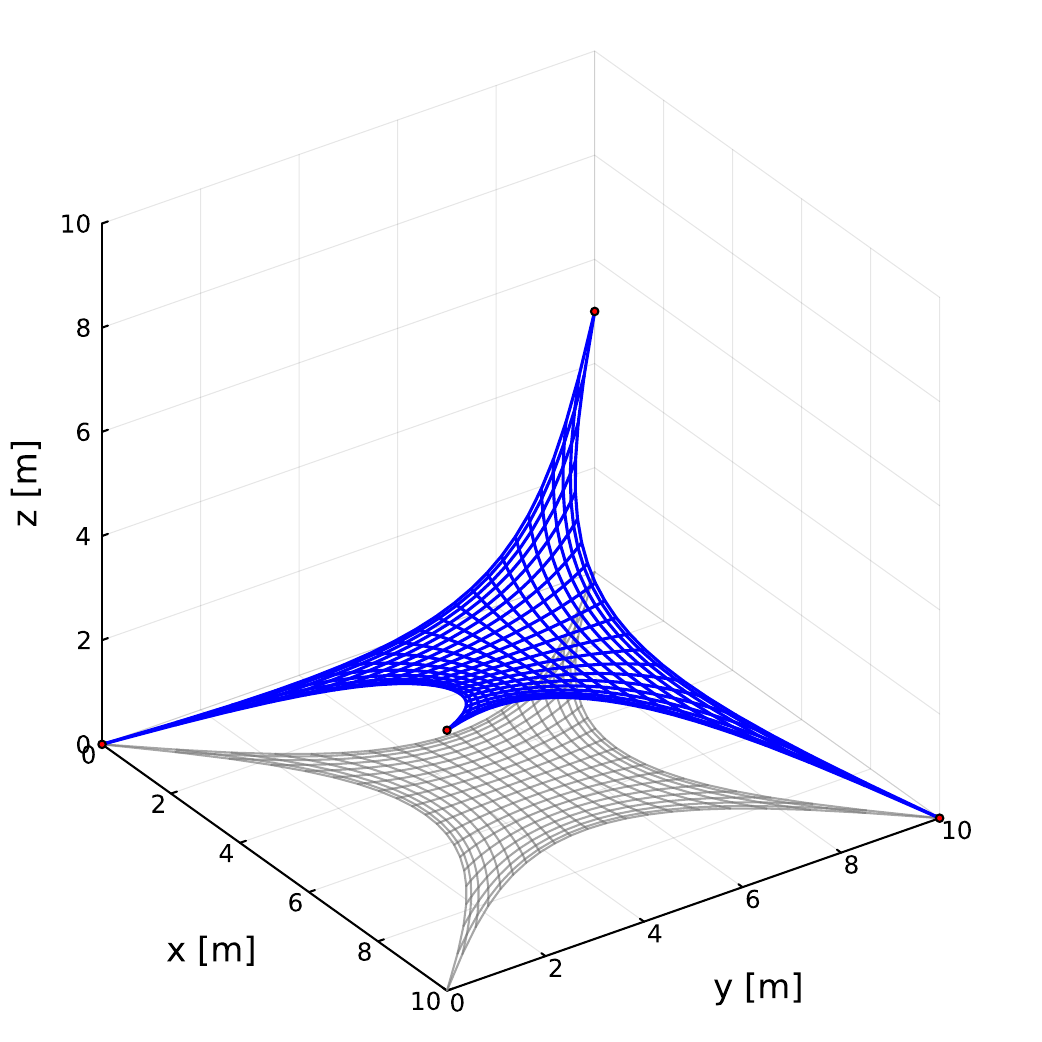}
        \caption{}
        \label{fig:ex1}
    \end{subfigure}
\hfill
    \begin{subfigure}{0.475\textwidth}
        \centering
        \includegraphics[width=\textwidth]{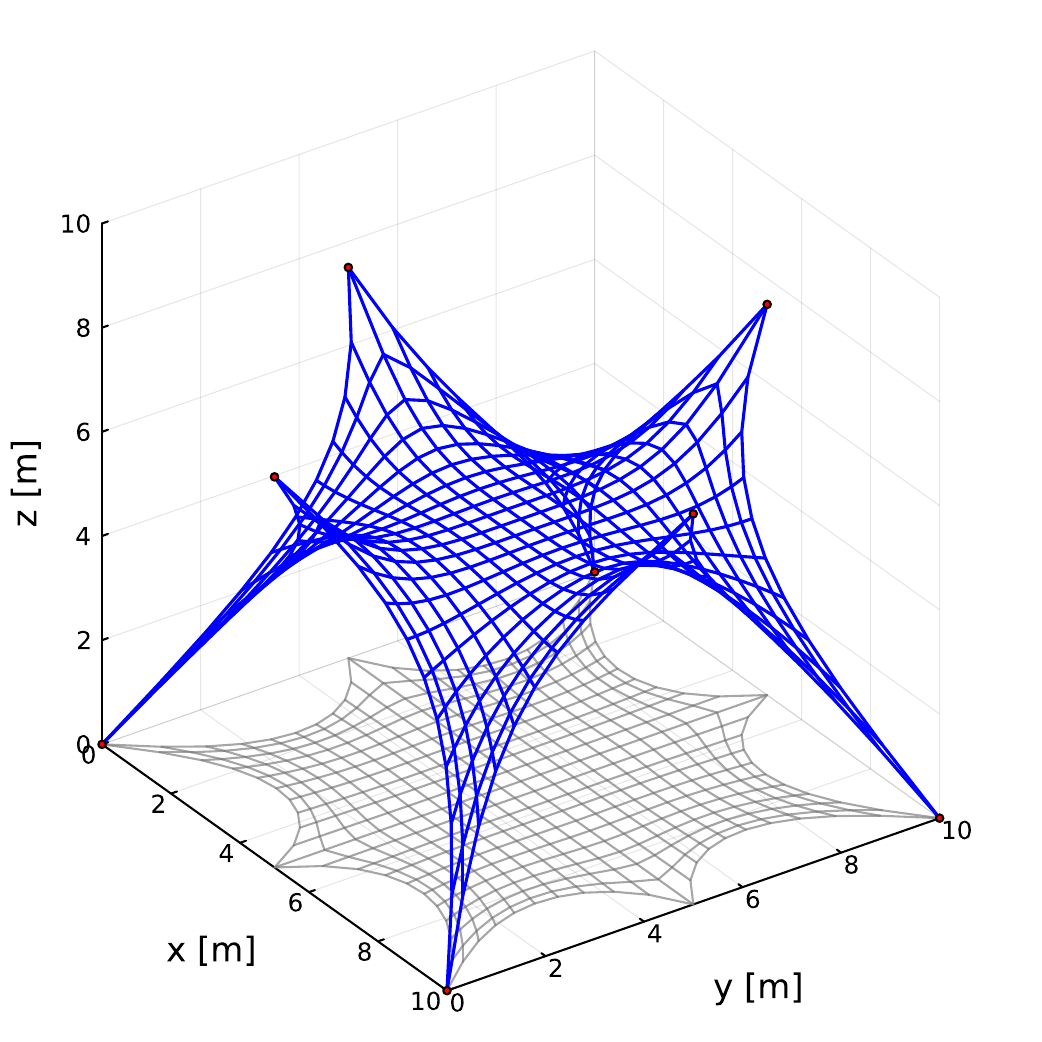}
        \caption{}
        \label{fig:ex2}
    \end{subfigure}
    \caption{Equilibrium configurations obtained with the FDM for two layouts with discrete fixed nodes: (a) a layout with two elevated fixed nodes and (b) a layout with four elevated fixed nodes. The gray grids show the horizontal projections of the equilibrated cable nets onto the $(x,y)$ plane, the blue grids show the equilibrated configurations, and the red markers indicate fixed nodes. In both cases, the boundary and internal elements have force densities of $2$ and $1$~N/m, respectively.}
    \label{fig:examples_fdm1}
\end{figure*}

\begin{figure*}[htbp!]
    \centering
    \begin{subfigure}{0.475\textwidth}
        \centering
        \includegraphics[width=\textwidth]{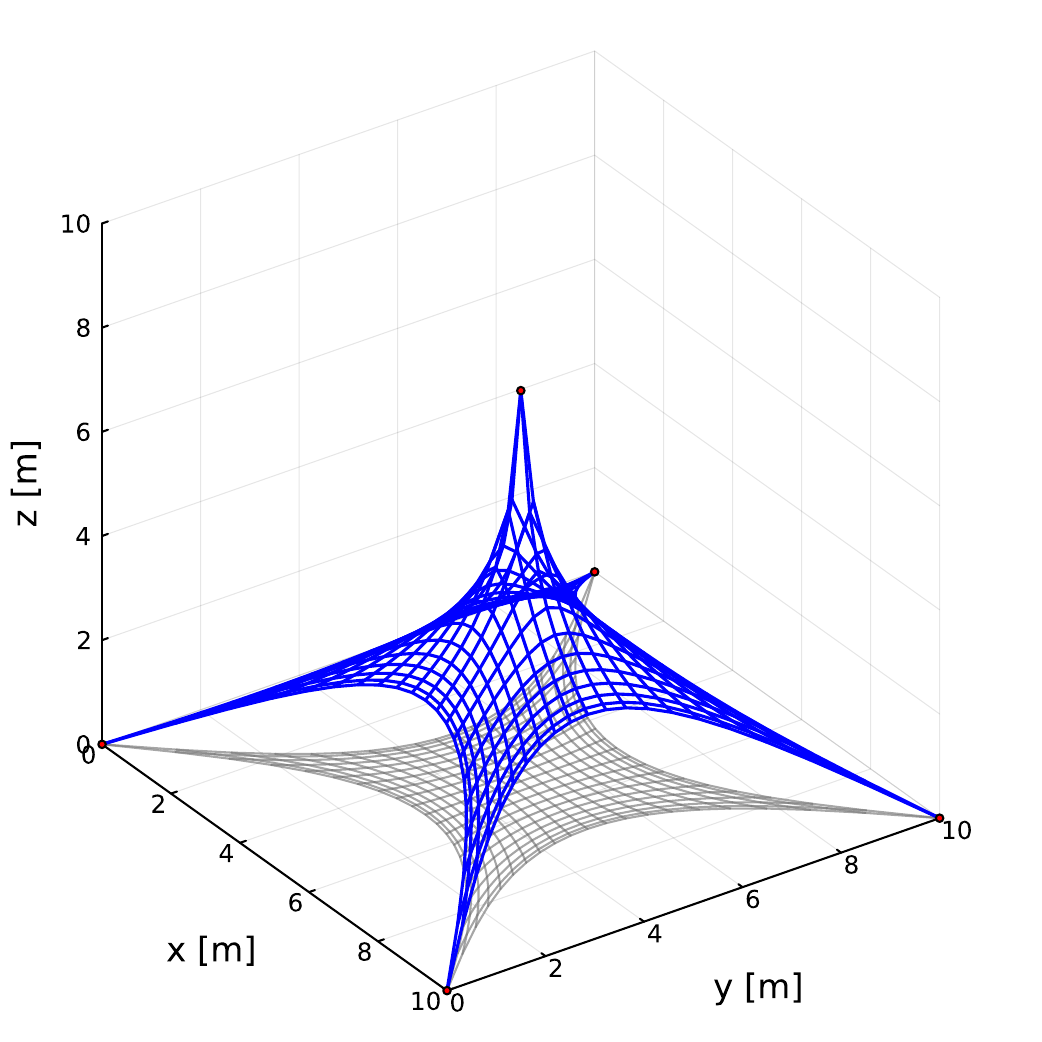}
        \caption{}
        \label{fig:ex3}
    \end{subfigure}
\hfill
    \begin{subfigure}{0.475\textwidth}
        \centering
        \includegraphics[width=\textwidth]{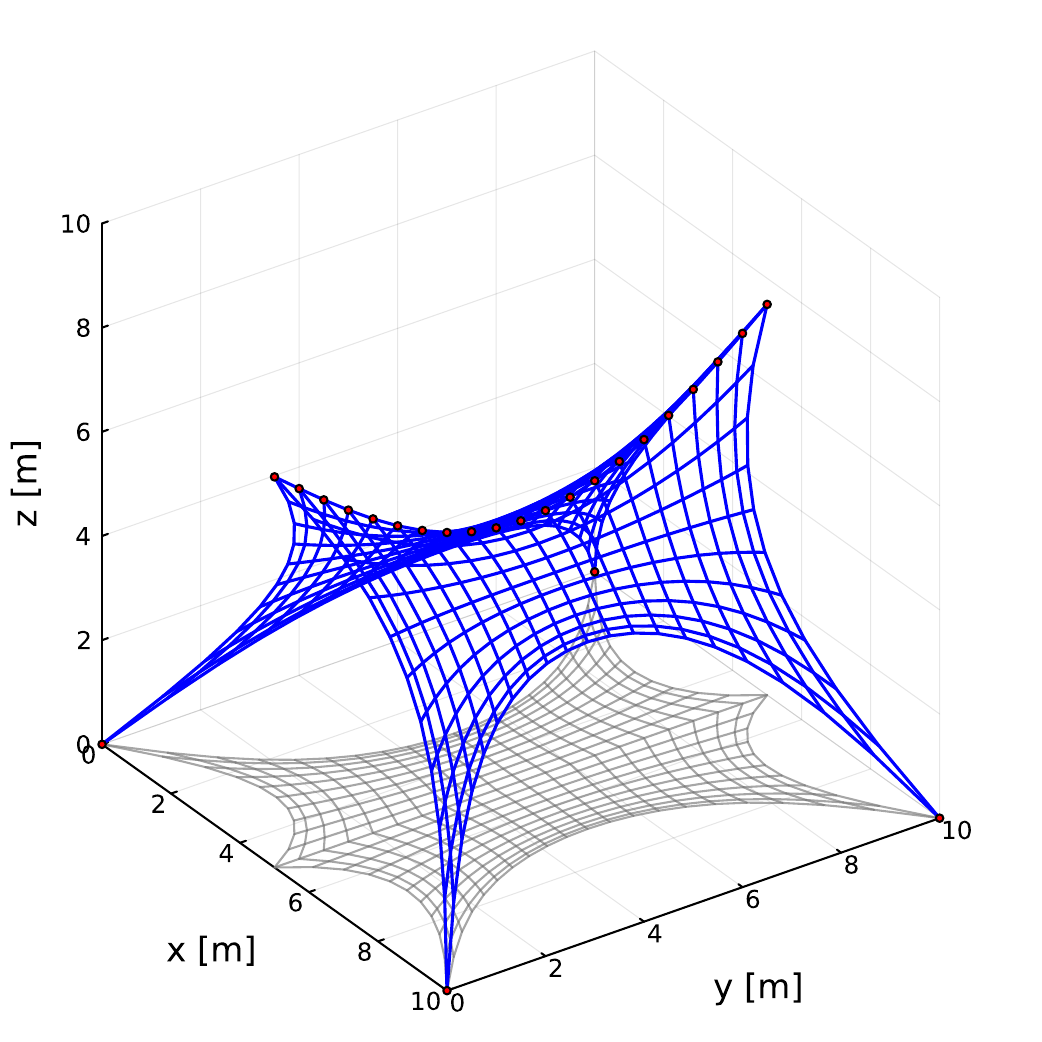}
        \caption{}
        \label{fig:ex4}
    \end{subfigure}

    \caption{Equilibrium configurations obtained with the FDM for two additional support layouts: (a) a layout with one elevated fixed node at the center and (b) a layout with fixed nodes distributed along an elevated boundary curve. The gray grids show the horizontal projections of the equilibrated cable nets onto the $(x,y)$ plane, the blue grids show the equilibrated configurations, and the red markers indicate fixed nodes. In both cases, the boundary and internal elements have force densities of $2$ and $1$~N/m, respectively.}
    \label{fig:examples_fdm2}
\end{figure*}

We consider a cable network.
At each node $i$ of the network, equilibrium between the external loads and internal axial forces in the cables is written as follows:
\begin{equation}\label{eq:eqeq1}
    \begin{split}
        & t_{i1}\frac{x_1-x_i}{l_{i1}} + \dots + t_{ik}\frac{x_k-x_i}{l_{ik}} + f_{xi} = 0 \\
        & t_{i1}\frac{y_1-y_i}{l_{i1}} + \dots + t_{ik}\frac{y_k-y_i}{l_{ik}} + f_{yi} = 0 \\
        & t_{i1}\frac{z_1-z_i}{l_{i1}} + \dots + t_{ik}\frac{z_k-z_i}{l_{ik}} + f_{zi} = 0, \\
    \end{split}
\end{equation}
where $k$ elements are assumed to connect node $i$ to $k$ surrounding nodes.
In the preceding equations, $t_{ij}$ and $l_{ij}$ are the tension and length of the cable connecting nodes $i$ and $j$, respectively, where $l_{ij} = \sqrt{(x_i-x_j)^2 + (y_i-y_j)^2 + (z_i-z_j)^2}$.
Before proceeding with the matrix formulation of the equilibrium equations of the cable network, we define the following vectors:
\begin{enumerate}
    \item $\mathbf{x}$, $\mathbf{y}$, and $\mathbf{z}$ are the vectors containing the nodal coordinates of the cable network. Their dimensions are $[n \times 1]$. The nodes are ordered such that the free nodes are followed by the fixed nodes. Consequently, the coordinate vectors can be split into free and fixed node coordinate vectors: $\mathbf{x}_f$, $\mathbf{y}_f$, and $\mathbf{z}_f$ for the free nodes, with dimensions $[n_f\times 1]$, and $\mathbf{x}_c$, $\mathbf{y}_c$, and $\mathbf{z}_c$ for the constrained or fixed nodes, with dimensions $[n_c\times 1]$;
    \item  $\mathbf{f}_x$, $\mathbf{f}_y$, and $\mathbf{f}_z$ are the load vectors associated with the $x$, $y$, and $z$ coordinate axes. Their dimensions are $[n\times 1]$;
    \item $\mathbf{t}$ and $\mathbf{l}$ are the vectors that collect the axial forces and lengths of the structural elements. Their dimensions are $[m\times 1]$. 
\end{enumerate}

Another important ingredient for writing the equilibrium equations in Eq.~\eqref{eq:eqeq1} in matrix form is the connectivity matrix $\mathbf{C}$, whose dimensions are $m\times n$. Each row of $\mathbf{C}$ is associated with one element of the cable network.
If element $i$ connects nodes $j$ and $k$, the nonzero entries in row $i$ of $\mathbf{C}$ are $C_{ij}=-1$ and $C_{ik}=1$. It follows that
\begin{equation}
    \begin{split}
        & \Delta \mathbf{x} = \mathbf{C}\,\mathbf{x};\\
        & \Delta \mathbf{y} = \mathbf{C}\,\mathbf{y};\\
        & \Delta \mathbf{z} = \mathbf{C}\,\mathbf{z}.
    \end{split}
\end{equation}
Because of the node ordering, the coordinate vectors can be split into complementary vectors associated with free ``f'' and constrained ``c'' nodes. For example,
\[
\mathbf{x}
=
\begin{bmatrix}
\mathbf{x}_f \\
\mathbf{x}_c
\end{bmatrix},
\]
with analogous expressions for the coordinate vectors $\mathbf{y}$ and $\mathbf{z}$. This partition yields
\begin{equation}\label{eq:dofsplit}
    \begin{split}
        & \Delta \mathbf{x} = \mathbf{C}_f\,\mathbf{x}_f + \mathbf{C}_c\,\mathbf{x}_c\\
        & \Delta \mathbf{y} = \mathbf{C}_f\,\mathbf{y}_f + \mathbf{C}_c\,\mathbf{y}_c\\
        & \Delta \mathbf{z} = \mathbf{C}_f\,\mathbf{z}_f + \mathbf{C}_c\,\mathbf{z}_c.
    \end{split}
\end{equation}

The equilibrium equations \eqref{eq:eqeq1} can then be written in matrix form as follows:
\begin{equation}\label{eq:eqeq2}
        \begin{bmatrix}
        \mathbf{C}^T_f \, \mathbf{\Delta X} \, \mathbf{L}^{-1}\\
        \mathbf{C}^T_f \, \mathbf{\Delta Y} \, \mathbf{L}^{-1}\\
        \mathbf{C}^T_f \, \mathbf{\Delta Z} \, \mathbf{L}^{-1}\\
        \end{bmatrix}
        \mathbf{t} = 
        \begin{bmatrix}
            \mathbf{f}_x\\
            \mathbf{f}_y\\
            \mathbf{f}_z\\
        \end{bmatrix} 
        \equiv
        \mathbf{A}\, \mathbf{t} = \mathbf{f}.
\end{equation}
In Eq.~\eqref{eq:eqeq2}, $\mathbf{\Delta X}$ is a diagonal matrix containing the entries of the vector $\mathbf{\Delta x}$, that is, $\mathbf{\Delta X}=\text{diag}(\mathbf{\Delta x})$. Analogous definitions are used for $\mathbf{\Delta Y}$, $\mathbf{\Delta Z}$, and $\mathbf{L}$ with respect to the vectors $\mathbf{\Delta y}$, $\mathbf{\Delta z}$, and $\mathbf{l}$.
The system of equations in Eq.~\eqref{eq:eqeq2} is nonlinear because the element lengths depend on the nodal coordinates of the cable network.

At this point, the concept of force density can be introduced. The scalar quantity $q_i=t_i/l_i$ is defined for each element $i$ in the cable network. The force densities $q_i$ are collected in the vector
\begin{equation}\label{eq:forcedens}
    \mathbf{q} = \mathbf{L}^{-1}\mathbf{t}.
\end{equation}
With the definition of Eq.~\eqref{eq:forcedens}, the system of equations \eqref{eq:eqeq2} becomes linear:
\begin{equation}\label{eq:force-density-balance}
    \begin{split}
        & \mathbf{C}^T_f \, \mathbf{\Delta X} \, \mathbf{q} = \mathbf{f}_x\\
        & \mathbf{C}^T_f \, \mathbf{\Delta Y} \, \mathbf{q} = \mathbf{f}_y\\
        & \mathbf{C}^T_f \, \mathbf{\Delta Z} \, \mathbf{q} = \mathbf{f}_z.\\
    \end{split}
\end{equation}
Defining the diagonal matrix $\mathbf{Q} = \text{diag}(\mathbf{q})$ leads to the following equalities:
\begin{equation}
    \begin{split}
    & \mathbf{\Delta X} \, \mathbf{q} = \mathbf{Q} \, \mathbf{\Delta x}\\
    & \mathbf{\Delta Y} \, \mathbf{q} = \mathbf{Q} \, \mathbf{\Delta y}\\
    & \mathbf{\Delta Z} \, \mathbf{q} = \mathbf{Q} \, \mathbf{\Delta z},\\
    \end{split}
\end{equation}
and thus Eq.~\eqref{eq:force-density-balance} becomes
\begin{equation}\label{eq:coordinate-equilibrium}
    \begin{split}
        & \mathbf{C}^T_f \, \mathbf{Q} \, \mathbf{\Delta x} = \mathbf{f}_x\\
        & \mathbf{C}^T_f \, \mathbf{Q} \, \mathbf{\Delta y} = \mathbf{f}_y\\
        & \mathbf{C}^T_f \, \mathbf{Q} \, \mathbf{\Delta z} = \mathbf{f}_z.\\
    \end{split}
\end{equation}
The vectors $\mathbf{\Delta x}$, $\mathbf{\Delta y}$, and $\mathbf{\Delta z}$ can be replaced by the expressions given in Eq.~\eqref{eq:dofsplit}, leading to
\begin{equation}\label{eq:partitioned-equilibrium}
    \begin{split}
        & \left(\mathbf{C}^T_f \, \mathbf{Q} \, \mathbf{C}_f\right) \, \mathbf{x}_f + \left(\mathbf{C}^T_f \, \mathbf{Q} \, \mathbf{C}_c\right) \, \mathbf{x}_c = \mathbf{f}_x\\
        & \left(\mathbf{C}^T_f \, \mathbf{Q} \, \mathbf{C}_f\right) \, \mathbf{y}_f + \left(\mathbf{C}^T_f \, \mathbf{Q} \, \mathbf{C}_c\right) \, \mathbf{y}_c = \mathbf{f}_y\\
        & \left(\mathbf{C}^T_f \, \mathbf{Q} \, \mathbf{C}_f\right) \, \mathbf{z}_f + \left(\mathbf{C}^T_f \, \mathbf{Q} \, \mathbf{C}_c\right) \, \mathbf{z}_c = \mathbf{f}_z.\\
    \end{split}
\end{equation}
Equation~\eqref{eq:partitioned-equilibrium} can be simplified by defining $\mathbf{D}_f=\mathbf{C}^T_f \, \mathbf{Q} \, \mathbf{C}_f$ and $\mathbf{D}_c=\mathbf{C}^T_f \, \mathbf{Q} \, \mathbf{C}_c$, thus obtaining
\begin{equation}\label{eq:fdm-linear-system}
    \begin{split}
        & \mathbf{D}_f \, \mathbf{x}_f = \mathbf{f}_x - \mathbf{D}_c \, \mathbf{x}_c \\
        & \mathbf{D}_f \, \mathbf{y}_f = \mathbf{f}_y - \mathbf{D}_c \, \mathbf{y}_c \\
        & \mathbf{D}_f \, \mathbf{z}_f = \mathbf{f}_z - \mathbf{D}_c \, \mathbf{z}_c . \\
    \end{split}
\end{equation}
The linear FDM is expressed by Eq.~\eqref{eq:fdm-linear-system}: once the force densities in all elements are prescribed, the coordinates of the free nodes are obtained as
\begin{equation}\label{eq:fdm-solution}
    \begin{split}
        & \mathbf{x}_f = \mathbf{D}_f^{-1}\left(\mathbf{f}_x - \mathbf{D}_c \, \mathbf{x}_c\right) \\
        & \mathbf{y}_f = \mathbf{D}_f^{-1}\left(\mathbf{f}_y - \mathbf{D}_c \, \mathbf{y}_c\right) \\
        & \mathbf{z}_f = \mathbf{D}_f^{-1}\left(\mathbf{f}_z - \mathbf{D}_c \, \mathbf{z}_c\right) .\\
    \end{split}
\end{equation}
Consequently, Eq.~\eqref{eq:fdm-solution} provides the equilibrated form, or configuration, of a cable structure once the external loads and element force densities are defined.

Examples are shown in Figs.~\ref{fig:examples_fdm1} and~\ref{fig:examples_fdm2}. They use a grid of $21\times21$ equally spaced nodes over an area of $10\times10$~m$^2$. The force densities must be prescribed to compute the equilibrated form of the cable network. In all examples of Figs.~\ref{fig:examples_fdm1} and~\ref{fig:examples_fdm2}, the boundary and internal elements are assigned force densities of $2$ and $1$~N/m, respectively. The red markers indicate nodes whose coordinates are prescribed and remain fixed.

\subsection{Nonlinear FDM}
\label{subsec:nonlinfdm}

In his seminal paper, \citet{Schek1974} also introduced a nonlinear extension of the FDM. The nonlinearity does not arise from the equilibrium equations themselves, which remain linear in the nodal coordinates for prescribed force densities. Instead, it results from imposing additional conditions that restrict the otherwise free choice of the force density vector. Schek considered prescribed distances between nodes, prescribed forces in selected members, prescribed unstressed member lengths, and combinations of these conditions. The number of nonlinear equations is therefore equal to the number of additional conditions and is independent of the number of free nodes in the network.
Let the additional conditions be collected in the vector valued function
\begin{equation}\label{eq:nlfdm-constraints}
    \mathbf{g}(\mathbf{r},\mathbf{q})=\mathbf{0},
\end{equation}
where $\mathbf{r}=[\mathbf{x}; \mathbf{y}; \mathbf{z}]$ collects the nodal coordinates and $\mathbf{q}$ contains the force densities. Because the equilibrium coordinates are obtained from the linear FDM as functions of the force densities, $\mathbf{r}=\mathbf{r}(\mathbf{q})$, Eq.~\eqref{eq:nlfdm-constraints} can be reduced to a system involving only $\mathbf{q}$:
\begin{equation}\label{eq:nlfdm-reduced}
    \mathbf{g}^*(\mathbf{q})
    =\mathbf{g}\bigl(\mathbf{r}(\mathbf{q}),\mathbf{q}\bigr)
    =\mathbf{0}.
\end{equation}
Starting from an equilibrium configuration obtained with the linear method, Schek solved this generally nonlinear system iteratively. At iteration $k$, the additional conditions are linearized as
\begin{equation}\label{eq:nlfdm-linearization}
    \mathbf{J}_g(\mathbf{q}^{(k)})\,\Delta\mathbf{q}
    =-\mathbf{g}^*(\mathbf{q}^{(k)}),
\end{equation}
where $\mathbf{J}_g=\partial\mathbf{g}^*/\partial\mathbf{q}$ is the Jacobian of the reduced constraint vector. The force densities are then updated according to $\mathbf{q}^{(k+1)}=\mathbf{q}^{(k)}+\Delta\mathbf{q}$. Because the number of additional conditions is often smaller than the number of members, Eq.~\eqref{eq:nlfdm-linearization} is generally underdetermined. Schek selected a unique correction by minimizing the Euclidean norm of $\Delta\mathbf{q}$. This minimum norm criterion selects, at each iteration, the smallest change in the force densities that satisfies the linearized additional conditions. The iterative procedure starts from an initial force density vector $\mathbf{q}_0$ and its associated equilibrium nodal coordinates $\mathbf{r}_0=[\mathbf{x}_0;\mathbf{y}_0;\mathbf{z}_0]$, obtained using the linear FDM. Schek also proposed damped and modified damped least squares variants to improve convergence when the required changes in the force densities or the network geometry are large. The interested reader is referred to \citet{Schek1974} for further details. The Extended FDM proposed by \citet{malerba2012extended} provides a further development of this class of approaches by allowing prescribed conditions on support reactions and mixed systems of cables and struts.

To solve the nonlinear FDM problem, we propose a novel constrained optimization formulation in which the nodal coordinates and force densities are treated as independent variables. In the proposed approach, $\mathbf{q}_0$ is an initial estimate of the force density vector. Solving the linear FDM with $\mathbf{q}_0$ produces the corresponding initial equilibrium form, described by the nodal coordinate vectors $\mathbf{x}_0$, $\mathbf{y}_0$, and $\mathbf{z}_0$. The force density vector $\mathbf{q}_0$ and the associated coordinate vectors are then used to initialize the nonlinear FDM optimization problem. We present the formulation for a specific case of the nonlinear constraints discussed by Schek, in which selected member lengths are prescribed while the optimized force densities are kept as close as possible to $\mathbf{q}_0$. Other constraint types listed by Schek could be added or substituted, leading to corresponding modifications of the optimization problem. Nevertheless, the main idea and the general class of problem formulated and solved within the nonlinear FDM framework would remain unchanged. For the case considered here, the resulting problem reads as follows:
\begin{equation}\label{eq:nlfdm-nlp}
\begin{split}
\minimize_{\mathbf{q},\mathbf{x}_f,\mathbf{y}_f,\mathbf{z}_f}\quad
    & \left\|\mathbf{q}-\mathbf{q}_0\right\|_2^2 \\
\text{subject to}\quad
    & \mathbf{C}^T_f\mathbf{Q}
      (\mathbf{C}_f\mathbf{x}_f+\mathbf{C}_c\mathbf{x}_c)=\mathbf{f}_x,\\
    & \mathbf{C}^T_f\mathbf{Q}
      (\mathbf{C}_f\mathbf{y}_f+\mathbf{C}_c\mathbf{y}_c)=\mathbf{f}_y,\\
    & \mathbf{C}^T_f\mathbf{Q}
      (\mathbf{C}_f\mathbf{z}_f+\mathbf{C}_c\mathbf{z}_c)=\mathbf{f}_z,\\
    & \left\|\mathbf{r}_a-\mathbf{r}_b\right\|_2^2=l_{0,e}^2,
      \quad e=(a,b)\in\mathcal{E}_c,\\
    & q_e\geq q_{\min},\quad e=1,\ldots,m,
\end{split}
\end{equation}
where $\mathbf{r}_a=[x_a,y_a,z_a]^T$ is the position of node $a$, $\mathcal{E}_c$ is the set of members with prescribed lengths, and $l_{0,e}$ is the target length of member $e$. The lower bound $q_{\min}>0$ maintains a strictly positive force density in every cable and avoids members with zero force density. This direct formulation avoids the explicit derivation of the coordinate sensitivities required by Schek's reduced formulation and can be solved, for example, using a general gradient-based nonlinear programming algorithm with automatic differentiation of the functions' gradients.

\begin{figure}
    \centering
    \includegraphics[width=0.9\linewidth]{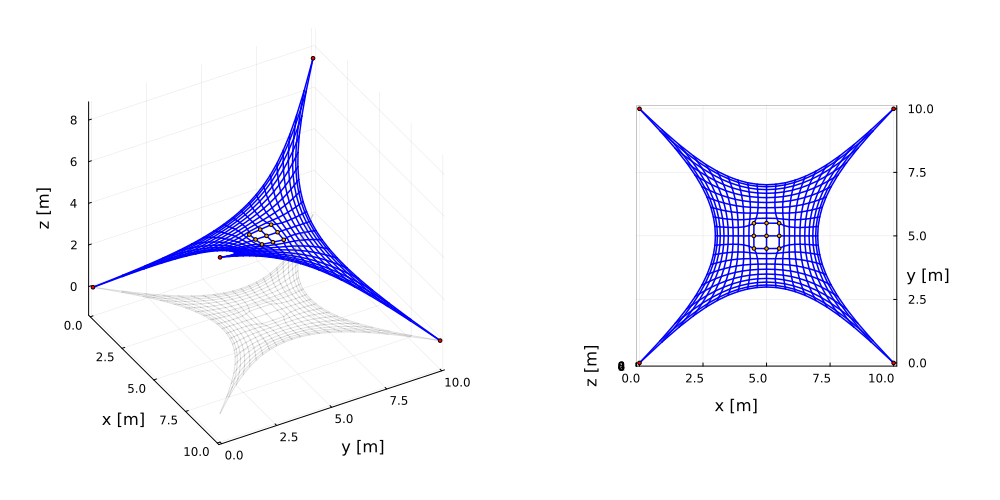}
    \caption{Equilibrium configuration obtained with the nonlinear FDM using the same boundary conditions and force density settings as in Fig.~\ref{fig:ex1}. Additional nonlinear constraints preserve the initial distances between the nodes highlighted in orange. The left and right panels show perspective and plan views, respectively. The gray grid shows the horizontal projection of the equilibrated cable net onto the $(x,y)$ plane, the blue grid shows the equilibrated configuration, and the red markers indicate fixed nodes.}
    \label{fig:nonlinfdm}
\end{figure}

The nonlinear formulation in Eq.~\eqref{eq:nlfdm-nlp} is illustrated using a $21\times21$ orthogonal cable net defined over a $10\times10$~m$^2$ square domain. The network contains $441$ nodes and $840$ horizontal and vertical members; diagonal members are not included. The four corner nodes are fixed at alternating elevations of $0$ and $7.5$~m. No external nodal loads are applied. The reference force densities are set to $q_{0b}=2$~N/m for the boundary members and $q_{0i}=1$~N/m for the internal members.
Additional geometric conditions are imposed on the $3\times3$ patch of nodes centered at $(x,y)=(5,5)$~m. Specifically, the lengths of its six horizontal and six vertical members are prescribed to remain equal to their initial value of $0.5$~m. Thus, $\mathcal{E}_c$ contains $12$ members and $l_{0,e}=0.5$~m for every $e\in\mathcal{E}_c$. The ordinary linear FDM is first solved using $\mathbf{q}_0$ to provide an initial equilibrium guess for the nodal coordinates. Problem~\eqref{eq:nlfdm-nlp} is then formulated in JuMP.jl \citep{dunning2017jump} and solved with IPOPT \citep{wachter2006implementation}; MadNLP can be used alternatively \citep{shin2021graph,shin2024accelerating}. In the numerical implementation, the force densities are constrained by $q_e\geq10^{-4}$~N/m. The solution obtained at the end of the optimization analysis is shown in Fid.~\ref{fig:nonlinfdm}. It represents the equilibrium cable net that satisfies all $12$ prescribed member length conditions while minimizing the change in force densities relative to $\mathbf{q}_0$. 

\section{Topology optimization of force densities}
\label{sec:topopt}

With the form finding approach discussed in Sec.~\ref{sec:fdm}, the equilibrated configuration of a cable structure can be computed for prescribed external loads, fixed nodes, and element force densities. The objective is now to determine a second equilibrium configuration that uses only a prescribed fraction of the structural volume of the original cable network. Removing cable elements and modifying force transmission between the network joints changes the equilibrium conditions at the free nodes and generally produces a different equilibrated form. The goal is therefore to reduce the structural volume of the original cable network by removing selected elements while obtaining an equilibrium configuration with an optimized topology that remains as close as possible to the original geometry.

\subsection{Integer problem formulation}

We define two equilibrated forms: the reference configuration and the configuration with optimized topology.
The reference configuration is defined by the nodal coordinates $\mathbf{r}_0=[\mathbf{x}_0; \mathbf{y}_0; \mathbf{z}_0]$, while the configuration with optimized topology is characterized by the nodal coordinates $\mathbf{r}=[\mathbf{x}; \mathbf{y}; \mathbf{z}]$. Both $\mathbf{r}_0$ and $\mathbf{r}$ are column vectors. For a given set of fixed loads, the configuration $\mathbf{r}_0$ is in equilibrium with the force density vector $\mathbf{q}_0$. Similarly, the configuration $\mathbf{r}$ with optimized topology is in equilibrium with $\mathbf{q}$.

We introduce the vector $\bm{\gamma}$, which contains the design optimization variables $\gamma_i \in \{0,1\}$ for $i=1,\dots,m$, one for each cable element of the reference structure. 
The force densities in the optimized cable system are then
\begin{equation}\label{eq:q_q0_gamma}
    q_i = \gamma_i \, q_{0,i}
\end{equation}
which in matrix notation reads $\mathbf{q} = \bm{\Gamma} \mathbf{q}_0$, where $\bm{\Gamma} = \text{diag}(\bm{\gamma})$. 
Eq.~\eqref{eq:q_q0_gamma} defines the force densities of the optimized network as an interpolation of those associated with the reference configuration. In this way, the optimized configuration retains the original force density pattern while allowing only a subset of cable elements to remain active, or ``survive'', at the end of the topology optimization process.
Essentially, the topology optimization process consists of finding a new equilibrated configuration, or form, $\mathbf{r}(\mathbf{q}(\bm{\gamma}))$ that is as close as possible to the reference form $\mathbf{r}_0(\mathbf{q}_0)$. Numerically, this is achieved by minimizing the error $\left\| \mathbf{r}(\mathbf{q}(\bm{\gamma})) - \mathbf{r}_0(\mathbf{q}_0) \right\|_2^2$.
In addition, the optimized force density distribution must satisfy equilibrium under the prescribed external loads. Therefore, the optimization problem is constrained to identify admissible force density topologies whose corresponding nodal configurations are in static equilibrium. This requirement is imposed through the following equality constraints:
\begin{equation}\label{eq:eqeq_fdm_constr}
    \begin{split}
        & \left(\mathbf{C}^T_f \, \mathbf{Q}(\bm{\gamma}) \, \mathbf{C}_f\right) \, \mathbf{x}_f + \left(\mathbf{C}^T_f \, \mathbf{Q}(\bm{\gamma}) \, \mathbf{C}_c\right) \, \mathbf{x}_c = \mathbf{f}_x\\
        & \left(\mathbf{C}^T_f \, \mathbf{Q}(\bm{\gamma}) \, \mathbf{C}_f\right) \, \mathbf{y}_f + \left(\mathbf{C}^T_f \, \mathbf{Q}(\bm{\gamma}) \, \mathbf{C}_c\right) \, \mathbf{y}_c = \mathbf{f}_y\\
        & \left(\mathbf{C}^T_f \, \mathbf{Q}(\bm{\gamma}) \, \mathbf{C}_f\right) \, \mathbf{z}_f + \left(\mathbf{C}^T_f \, \mathbf{Q}(\bm{\gamma}) \, \mathbf{C}_c\right) \, \mathbf{z}_c = \mathbf{f}_z.\\
    \end{split}
\end{equation}

Thus, the topology optimization problem for the force densities is formulated as follows:
\begin{equation}\label{eq:optprob1}
\begin{split}
\minimize_{\bm{\gamma},\mathbf{x},\mathbf{y},\mathbf{z}} \quad & J(\bm{\gamma},\mathbf{x},\mathbf{y},\mathbf{z})
= \left\|
\mathbf{r} - \mathbf{r}_0
\right\|_2^2 \\ 
\text{subject to } & \mathbf{q}(\bm{\gamma}) = \bm{\Gamma} \mathbf{q}_0\\
        & \mathbf{C}^T_f \, \mathbf{Q}(\bm{\gamma}) \, \left(\mathbf{C}_f \, \mathbf{x}_f +  \mathbf{C}_c \, \mathbf{x}_c \right) = \mathbf{f}_x\\
        & \mathbf{C}^T_f \, \mathbf{Q}(\bm{\gamma}) \, \left(\mathbf{C}_f \, \mathbf{y}_f +  \mathbf{C}_c \, \mathbf{y}_c \right) = \mathbf{f}_y\\
        & \mathbf{C}^T_f \, \mathbf{Q}(\bm{\gamma}) \, \left(\mathbf{C}_f \, \mathbf{z}_f +  \mathbf{C}_c \, \mathbf{z}_c \right) = \mathbf{f}_z.\\
        & \sum_{i=1}^m \gamma_i l_{0,i} \leq V^* \\
        & \sum_{i \in \mathcal{E}_a} \gamma_i \geq \gamma_L^{\min}, \; \forall \; a \in \mathcal{N}_L\\
        & \gamma_i \in \{0,1\} \text{ for } i=1,\dots,m \\
        & \gamma_i = 1 \; \forall \; i \in \mathcal{B}
\end{split}
\end{equation}
In Problem~\eqref{eq:optprob1}, $V^*$ is the permitted structural volume, defined as a fraction of the volume of the reference cable structure whose coordinates are collected in $\mathbf{r}_0$, and $m$ is the number of cable elements. In addition, $\mathcal{N}_L$ is the set of loaded joints at which external loads are applied, and $\mathcal{E}_a$ is the set of cable elements connected at one end to loaded joint $a$. The parameter $\gamma_L^{\min}$ defines the minimum number of cable elements that must be connected to a loaded joint and is set to $2$ here.
The topology optimization problem in Eq.~\eqref{eq:optprob1} can be classified as a mixed binary bilinear programming problem. The nodal coordinate variables $\mathbf{x}$, $\mathbf{y}$, and $\mathbf{z}$ are real, whereas the variables $\bm{\gamma}$ are binary.
The next section presents a continuous relaxation of Problem~\eqref{eq:optprob1} and numerical techniques that promote convergence toward nearly discrete force density topologies despite the continuous definition of the optimization variables.

\subsection{Continuous problem relaxation}
The first step in reformulating Problem~\eqref{eq:optprob1} as a continuous topology optimization problem is to relax the definition of the optimization variables $\bm{\gamma}$. These variables are allowed to vary continuously between zero and one: $\gamma_i \in [0,1]$.
To discourage convergence toward intermediate values of $\bm{\gamma}$, we adopt two numerical strategies that penalize such values.
First, the intermediate values of $\gamma_i$ are penalized using the SIMP\footnote{Solid Isotropic Material with Penalization.} interpolation technique, which is commonly used in structural topology optimization based on density variables \citep{bendsoe1999material}.
The SIMP scheme penalizes intermediate values of the optimization variables $\gamma_i$ through the force density interpolation in Eq.~\eqref{eq:q_q0_gamma}:
\begin{equation}\label{eq:q_q0_gamma2}
    q_i = q_{min} + \gamma_i^p \, (q_{0,i}-q_{min})
\end{equation}
where $p>1$ is the penalization exponent; for $p=1$, the expression reduces to a linear interpolation. In addition, $q_{min}$ is a small force density value, for example $10^{-6}$, used to avoid numerical problems when $\gamma_i=0$.
We also apply the SIMP interpolation to the constraint that defines the minimum number of elements connected to each loaded joint
\begin{equation}\label{eq:nodeelepernode}
    \sum_{i \in \mathcal{E}_a} \gamma_i^p \geq \gamma_L^{\min}, \; \forall \; a \in \mathcal{N}_L.
\end{equation}
At the same time, the optimization variables $\bm{\gamma}$ are not penalized in the volume constraint
\begin{equation}\label{eq:volconstr}
    \sum_{i=1}^m \gamma_i l_{0,i} \leq V^*.
\end{equation}
The SIMP interpolation adopted in Eqs.~\eqref{eq:q_q0_gamma2} and~\eqref{eq:nodeelepernode}, together with the linear structural volume measure defined by Eq.~\eqref{eq:volconstr}, inherently penalizes intermediate values of the design variables. Although the contribution of an element to the volume constraint in Eq.~\eqref{eq:volconstr} varies linearly with $\gamma_i$, its force density contribution according to Eq.~\eqref{eq:q_q0_gamma2} is reduced by the SIMP penalization. Consequently, for $0<\gamma_i<1$, an element consumes a relatively large fraction of the available structural volume while providing only a limited contribution to the equilibrium of the cable network. This mismatch makes intermediate density values inefficient and implicitly promotes design variables with values close to zero or one.

\begin{figure}
    \centering
    \includegraphics[width=0.5\linewidth]{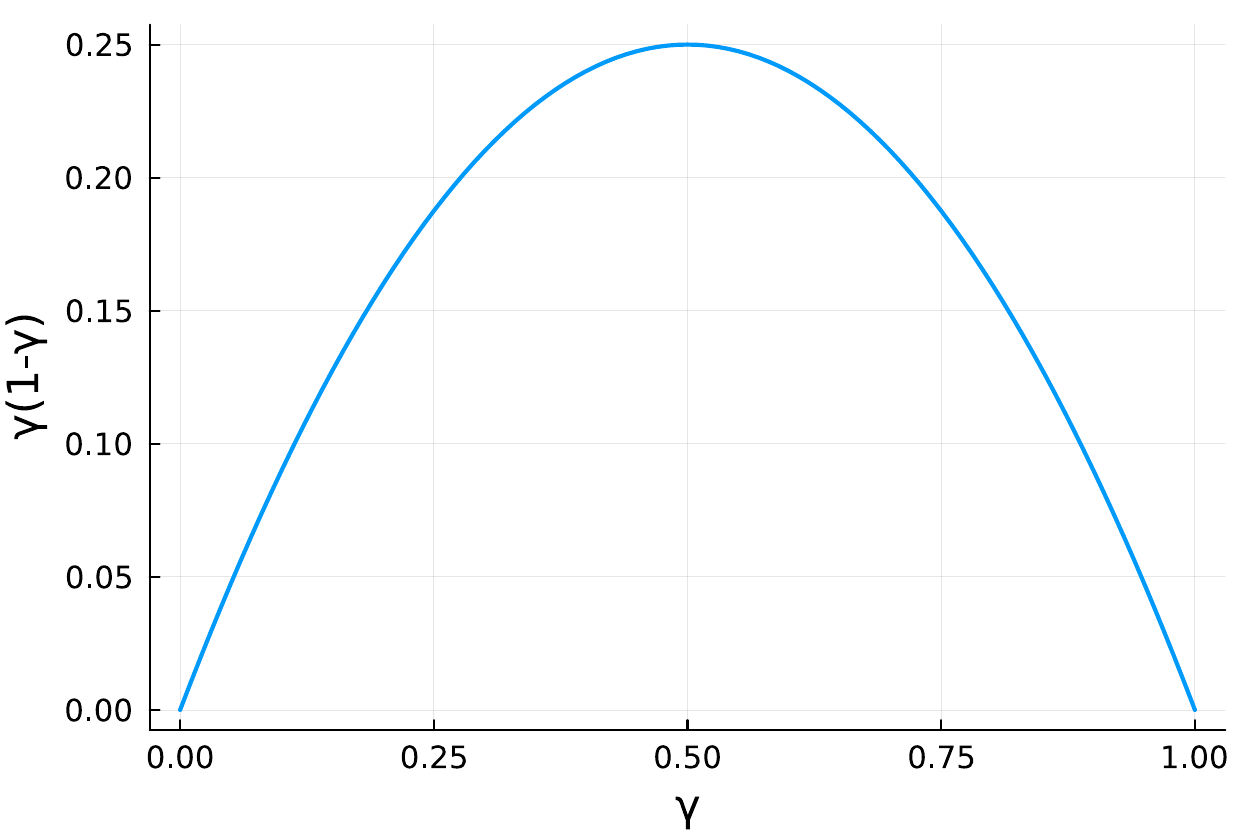}
    \caption{Binary penalization function $J_{\mathrm{bin}}(\gamma_i) = \gamma_i(1-\gamma_i)$ for a single design variable $\gamma_i \in [0,1]$. The function vanishes at $\gamma_i=0$ and $\gamma_i=1$ and penalizes the selection of intermediate density values.}
    \label{fig:jbin}
\end{figure}

To further promote convergence toward nearly binary values of the design variables, the following penalty term is added to the objective function:
\begin{equation}
\label{eq:objbin}
J_{\mathrm{bin}}(\bm{\gamma})=\sum_{i=1}^{m}\gamma_i \left(1-\gamma_i\right).
\end{equation}
The scalar function $\gamma_i(1-\gamma_i)$ is illustrated in Fig.~\ref{fig:jbin} for a single design variable $\gamma_i \in [0,1]$. This function vanishes at $\gamma_i=0$ and $\gamma_i=1$, and attains positive values for intermediate densities, with its maximum occurring at $\gamma_i=0.5$. Consequently, the penalty term in Eq.~\eqref{eq:objbin} is equal to zero only when every entry of $\bm{\gamma}$ is binary. When intermediate values are present, $J_{\mathrm{bin}}$ increases the objective function and therefore discourages non-discrete designs during the minimization process.

The final continuous optimization problem with the SIMP penalization scheme reads as follows:
\begin{equation}\label{eq:optprob2}
\begin{split}
\minimize_{\bm{\gamma},\mathbf{x},\mathbf{y},\mathbf{z}} \quad & J(\bm{\gamma},\mathbf{x},\mathbf{y},\mathbf{z})
= \left\|
\mathbf{r} - \mathbf{r}_0
\right\|_2^2 + \bm{\gamma}^T(\mathbf{1} - \bm{\gamma}) \\
\text{subject to } & \mathbf{q}(\bm{\gamma}) = \mathbf{q}_{min} +  \bm{\Gamma}^p \, (\mathbf{q}_0 - \mathbf{q}_{min})\\
        & \mathbf{C}^T_f \, \mathbf{Q}(\bm{\gamma}) \, \left(\mathbf{C}_f \, \mathbf{x}_f +  \mathbf{C}_c \, \mathbf{x}_c \right) = \mathbf{f}_x\\
        & \mathbf{C}^T_f \, \mathbf{Q}(\bm{\gamma}) \, \left(\mathbf{C}_f \, \mathbf{y}_f +  \mathbf{C}_c \, \mathbf{y}_c \right) = \mathbf{f}_y\\
        & \mathbf{C}^T_f \, \mathbf{Q}(\bm{\gamma}) \, \left(\mathbf{C}_f \, \mathbf{z}_f +  \mathbf{C}_c \, \mathbf{z}_c \right) = \mathbf{f}_z.\\
        & \sum_{i=1}^m \gamma_i l_{0,i} \leq V^* \\
        & \sum_{i \in \mathcal{E}_a} \gamma_i^p \geq \gamma_L^{\min}, \; \forall \; a \in \mathcal{N}_L\\
        & \gamma_i \in [0,1] \text{ for } i=1,\dots,m \\
        & \gamma_i = 1 \; \forall \; i \in \mathcal{B}.
\end{split}
\end{equation}

Problem~\eqref{eq:optprob2} is solved with an optimization algorithm based on first order information.
Specifically, MadNLP is used to solve the topology optimization problem. It is a nonlinear programming solver based on a filter line search interior point method similar to that used by IPOPT \citep{wachter2006implementation}.
The gradients of the objective and constraint functions are computed using automatic differentiation.
Further details on the software implementation are given in Sec.~\ref{subsec:softcons}.

\subsection{Software and computational considerations}
\label{subsec:softcons}
The proposed topology optimization approach was implemented in Julia \citep{bezanson2017julia}. The implementation covers both the FDM described in Sec.~\ref{sec:fdm}, which is used to generate the reference ground structures, and the topology optimization formulation presented in Sec.~\ref{sec:topopt}. The optimization problem is formulated using the JuMP.jl modeling language \citep{dunning2017jump} and solved with the interior point nonlinear programming solver MadNLP through the MadNLP.jl package \citep{shin2021graph,shin2024accelerating}. The required gradients of the objective and constraint functions are obtained through the automatic differentiation capabilities built into JuMP.jl.

\section{Numerical examples}
\label{sec:numex}

\begin{figure*}[htbp!]
\centering
\centering
\includegraphics[width=0.9\textwidth]{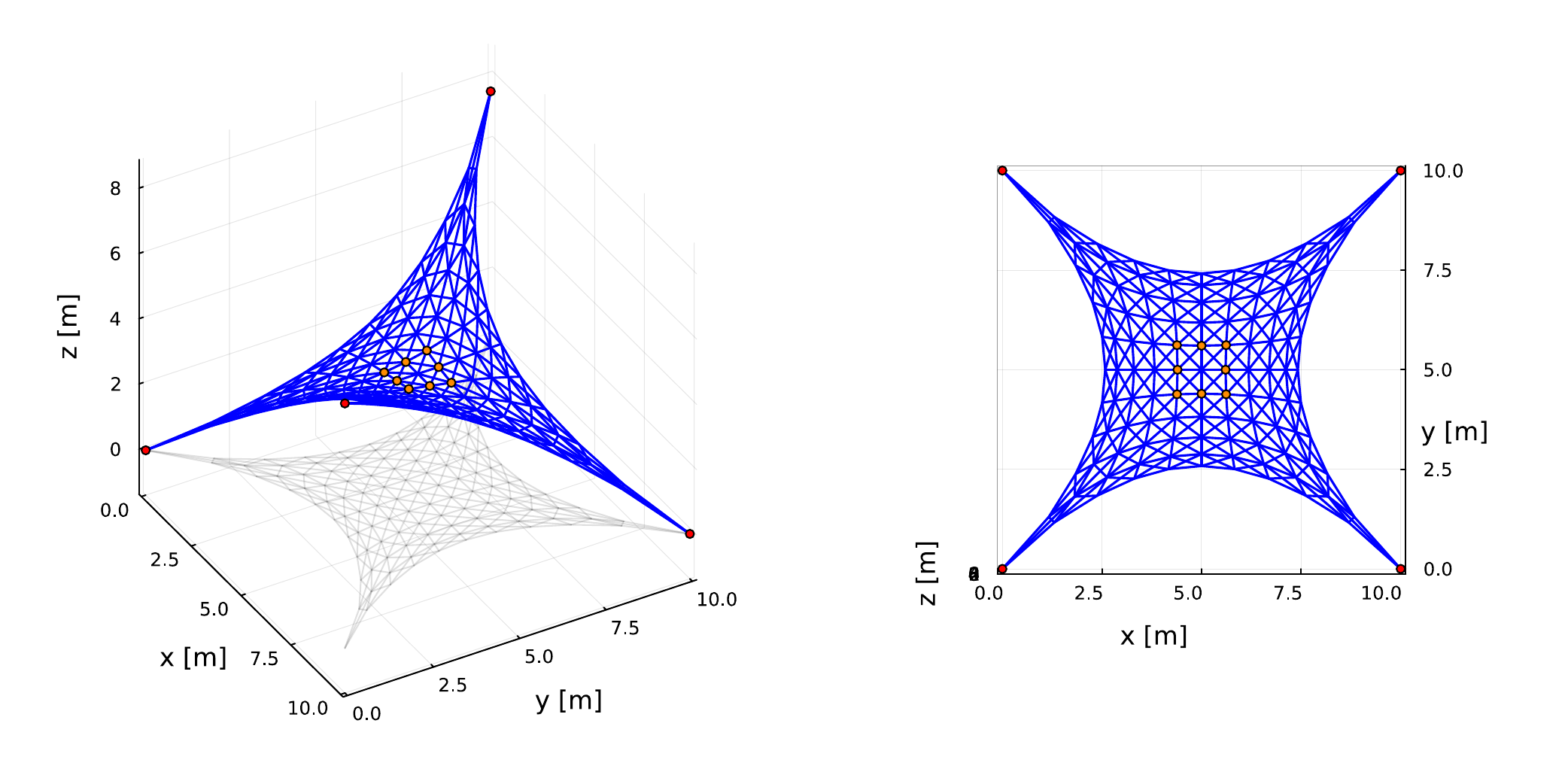}
\label{fig:ground21}
\caption{The first reference ground structure obtained by the linear FDM described in Sec.~\ref{subsec:linfdm}. The resulting force density vector $\mathbf{q}_0$ and equilibrium nodal coordinate vector $\mathbf{r}_0$ are subsequently used in Sec.~\ref{sec:numex} as reference quantities in the topology optimization analysis.}
\label{fig:groundrefs1}
\end{figure*}

\begin{figure*}[htbp!]
\centering
\includegraphics[width=0.9\textwidth]{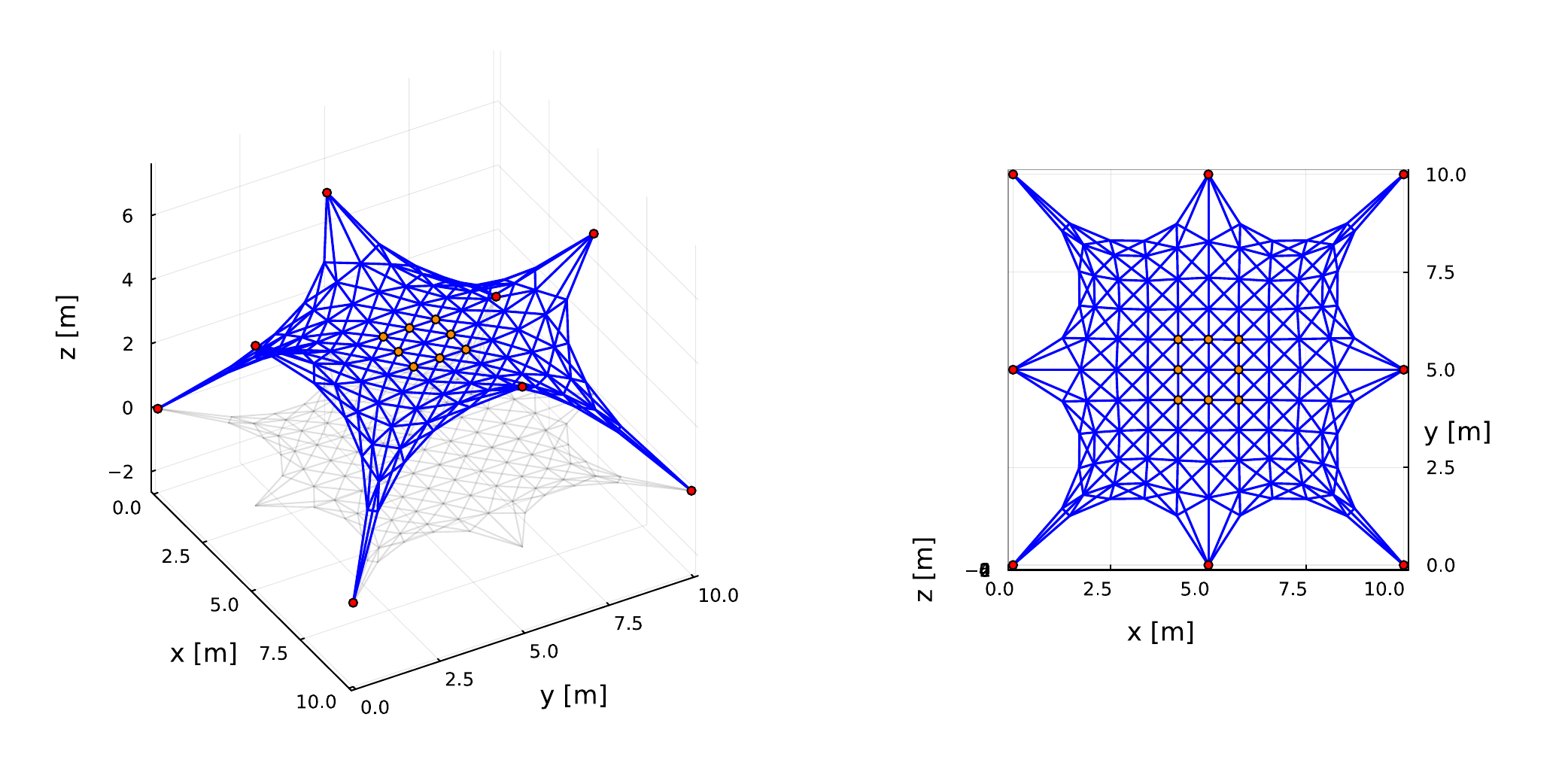}
\label{fig:ground22}
\caption{The second reference ground structure obtained by the linear FDM described in Sec.~\ref{subsec:linfdm}. The resulting force density vector $\mathbf{q}_0$ and equilibrium nodal coordinate vector $\mathbf{r}_0$ are subsequently used in Sec.~\ref{sec:numex} as reference quantities in the topology optimization analysis.}
\label{fig:groundrefs2}
\end{figure*}

\begin{figure*}[htbp!]
    \centering
        \centering
        \includegraphics[width=0.9\textwidth]{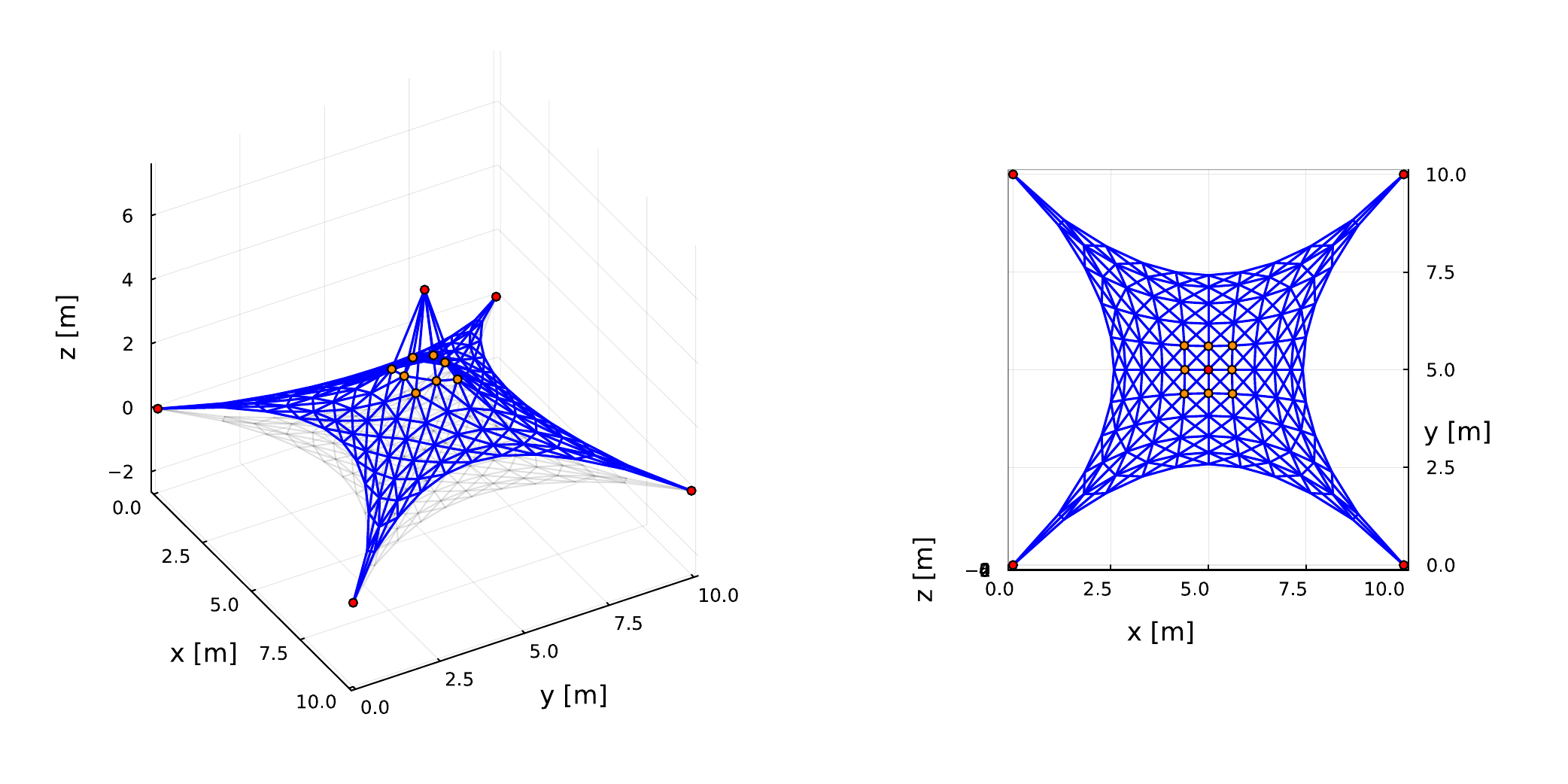}
    \caption{The third reference ground structure obtained by the linear FDM described in Sec.~\ref{subsec:linfdm}. The resulting force density vector $\mathbf{q}_0$ and equilibrium nodal coordinate vector $\mathbf{r}_0$ are subsequently used in Sec.~\ref{sec:numex} as reference quantities in the topology optimization analysis.}
    \label{fig:groundrefs3}
\end{figure*}

This section presents results for three design cases.
The cases are defined by different sets of boundary conditions and prescribed force densities in the reference ground structures, or cable nets.
The reference ground structure is defined by a regular network of horizontal and vertical cable elements aligned with the global coordinate axes $x$ and $y$. Each quadrilateral cell is also reinforced by two intersecting diagonal elements, resulting in a cable net configuration with X bracing.

The three reference ground structures obtained with the linear FDM (see Sec.~\ref{subsec:linfdm}) are shown in Figs.~\ref{fig:groundrefs1}, \ref{fig:groundrefs2} and \ref{fig:groundrefs3}. Fixed and loaded nodes are indicated in red and orange, respectively. All ground structures are defined over a square domain in the $(x,y)$ plane with dimensions $10 \times 10~\mathrm{m}^2$. Two nodal resolutions are considered: grids comprising $11 \times 11$ and $21 \times 21$ nodes.
For each case, the cable elements are divided into three groups: border, internal, and diagonal elements. Each group is assigned a different force density, namely $q_b$ for the border elements, $q_i$ for the internal elements, and $q_d$ for the internal diagonal elements.
The three cases are referred to as Example 1 (Fig.~\ref{fig:groundrefs1}), Example 2 (Fig.~\ref{fig:groundrefs2}), and Example 3 (Fig.~\ref{fig:groundrefs3}).
The force density values considered for the three examples are:
\begin{itemize}
    \item Example 1: $q_b=5~\mathrm{kN/m}$, $q_i=1~\mathrm{kN/m}$, $q_d=1~\mathrm{kN/m}$.
    \item Example 2: $q_b=3~\mathrm{kN/m}$, $q_i=2~\mathrm{kN/m}$, $q_d=1~\mathrm{kN/m}$.
    \item Example 3: $q_b=5~\mathrm{kN/m}$, $q_i=1~\mathrm{kN/m}$, $q_d=1~\mathrm{kN/m}$.
\end{itemize}
For all cases, a total vertical load of magnitude $1$~kN is distributed uniformly among the ring of nodes surrounding the central node. This fixed loading pattern is retained throughout the numerical examples for simplicity and to enable consistent comparison among the different ground structures. The load applied to each loaded node, shown in orange in Figs.~\ref{fig:groundrefs1}, \ref{fig:groundrefs2}, and~\ref{fig:groundrefs3}, is therefore $-0.125$~kN.
The support conditions, loading pattern, force density assignment, and ground structure connectivity used in these examples are selected for illustrative purposes. Different choices could be adopted to represent alternative design requirements. However, these modeling parameters directly influence the resulting equilibrium configurations and topology optimization outcomes.
Unless otherwise specified, the parameters $V^*$ and $\gamma_{min}^L$ are set to $0.4$ and $2$, respectively, in the following examples.

\begin{figure*}[htbp]
    \centering
    \begin{subfigure}{0.9\textwidth}
        \centering
        \includegraphics[width=\textwidth]{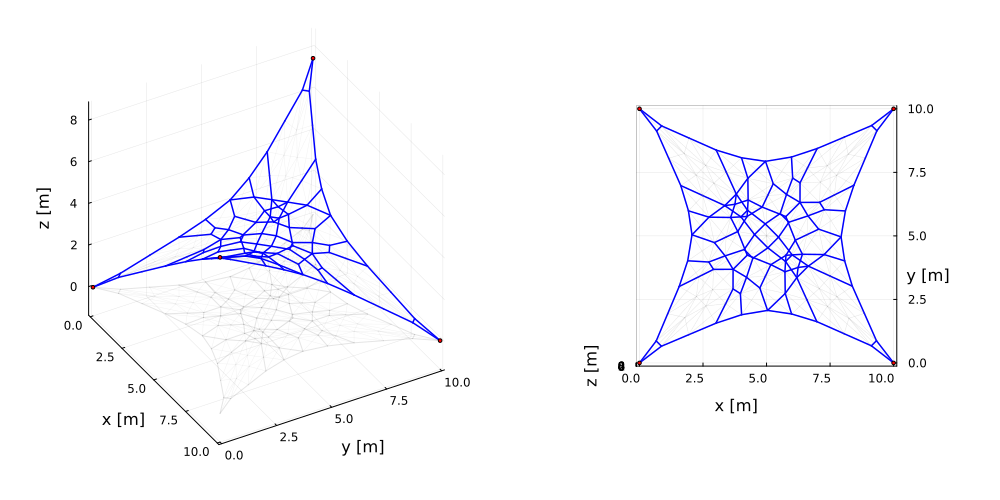}
        \caption{Results for the $11\times11$ reference cable network.}
        \label{fig:resex11}
    \end{subfigure}
\vfill
    \begin{subfigure}{0.9\textwidth}
        \centering
        \includegraphics[width=\textwidth]{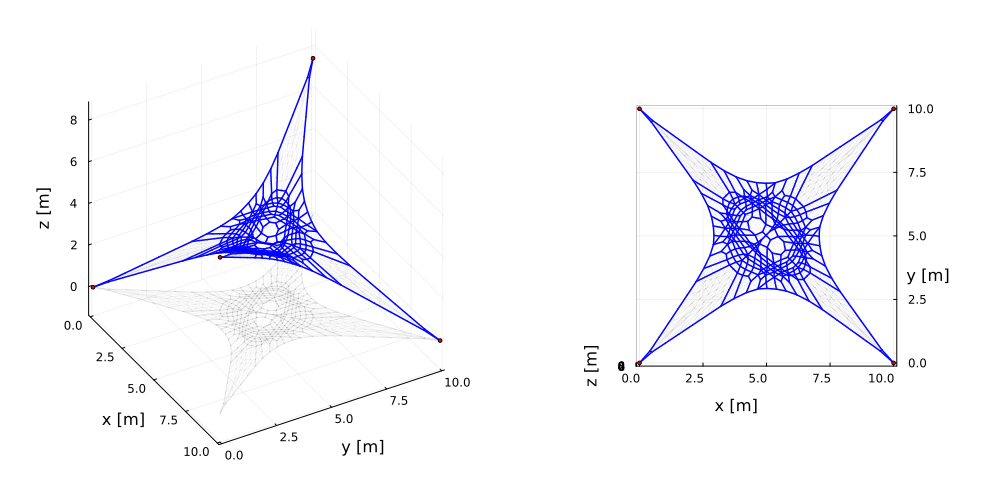}
        \caption{Results for the $21\times21$ reference cable network.}
        \label{fig:resex12}
    \end{subfigure}

    \caption{Topology optimization results for Example 1, whose reference ground structure is shown in Fig.~\ref{fig:groundrefs1}. The prescribed force densities are $q_b=5~\mathrm{kN/m}$, $q_i=1~\mathrm{kN/m}$, and $q_d=1~\mathrm{kN/m}$.}
    \label{fig:optres1}
\end{figure*}

\begin{figure*}[htbp]
    \centering
    \begin{subfigure}{0.9\textwidth}
        \centering
        \includegraphics[width=\textwidth]{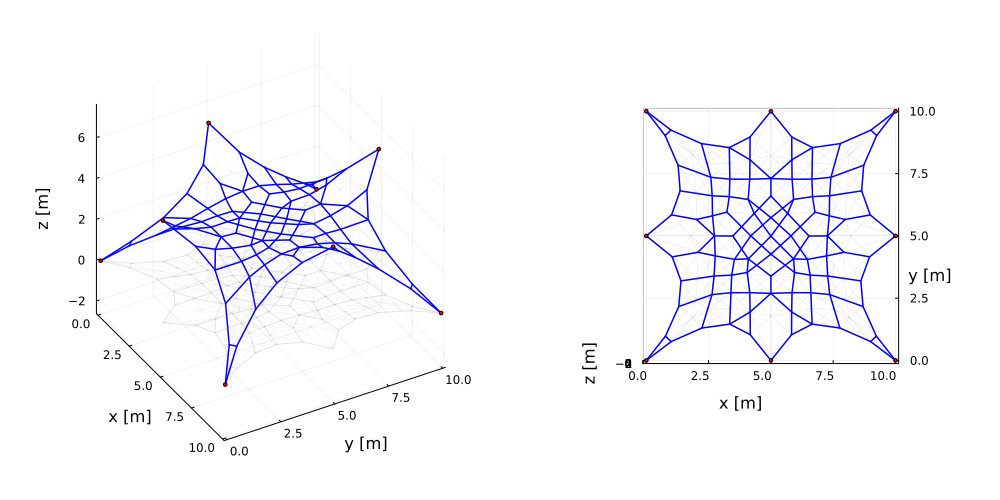}
        \caption{Results for the $11\times11$ reference cable network.}
        \label{fig:resex21}
    \end{subfigure}
\vfill
    \begin{subfigure}{0.9\textwidth}
        \centering
        \includegraphics[width=\textwidth]{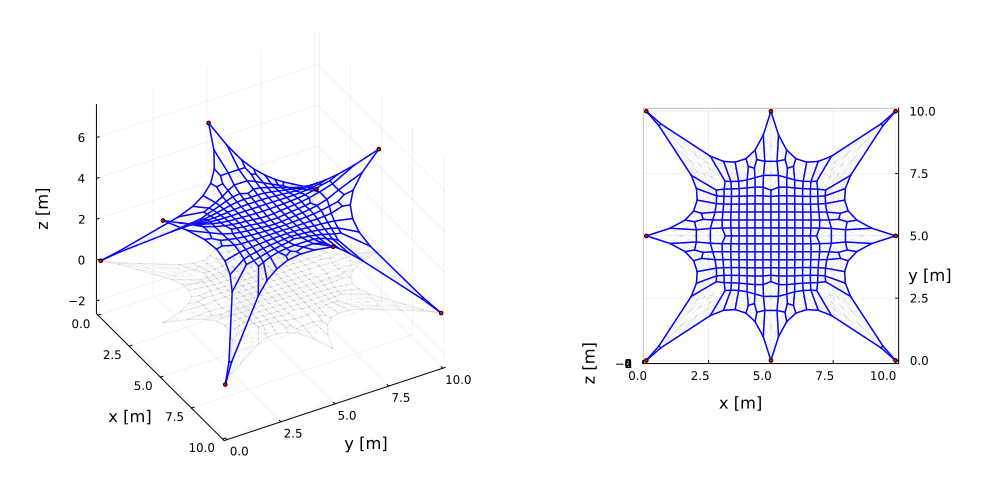}
        \caption{Results for the $21\times21$ reference cable network.}
        \label{fig:resex22}
    \end{subfigure}

    \caption{Topology optimization results for Example 2, whose reference ground structure is shown in Fig.~\ref{fig:groundrefs2}. The prescribed force densities are $q_b=3~\mathrm{kN/m}$, $q_i=2~\mathrm{kN/m}$, and $q_d=1~\mathrm{kN/m}$.}
    \label{fig:optres2}
\end{figure*}

\begin{figure*}[htbp]
    \centering
    \begin{subfigure}{0.9\textwidth}
        \centering
        \includegraphics[width=\textwidth]{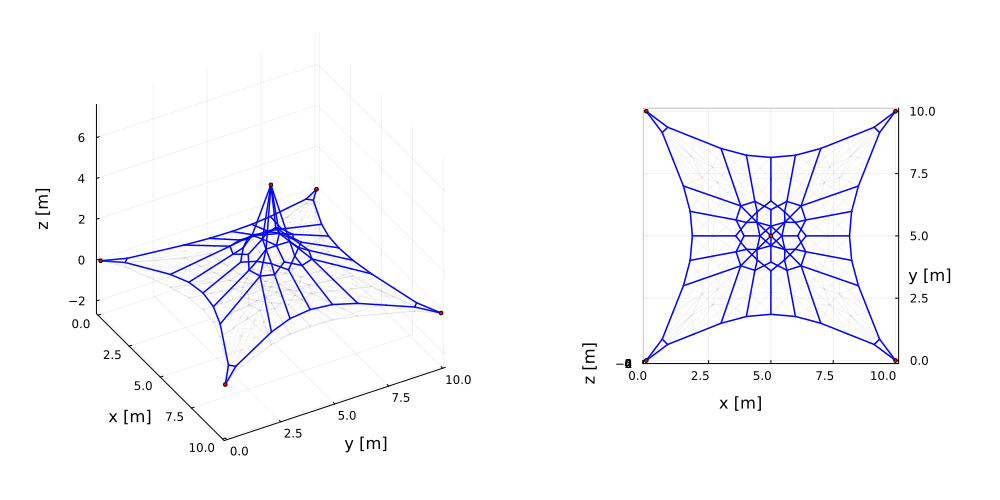}
        \caption{Results for the $11\times11$ reference cable network.}
        \label{fig:resex31}
    \end{subfigure}
\vfill
    \begin{subfigure}{0.9\textwidth}
        \centering
        \includegraphics[width=\textwidth]{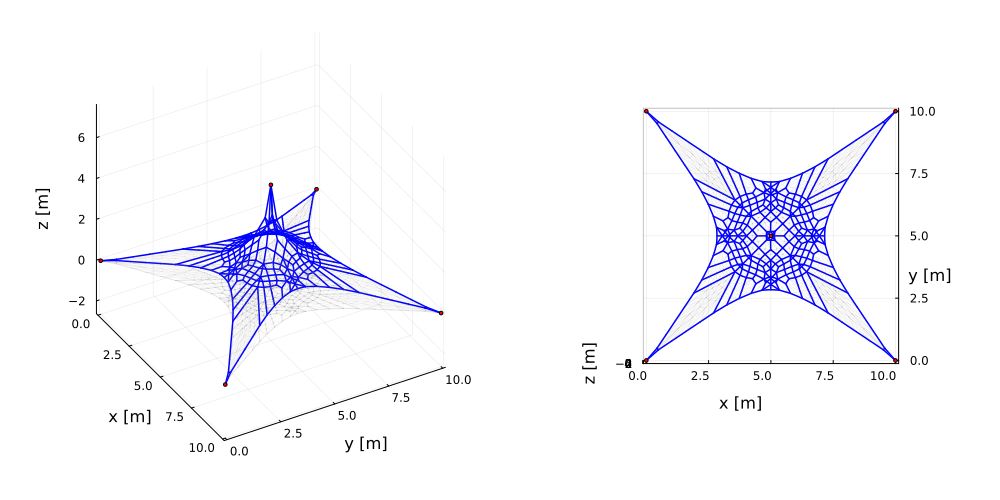}
        \caption{Results for the $21\times21$ reference cable network.}
        \label{fig:resex32}
    \end{subfigure}

    \caption{Topology optimization results for Example 3, whose reference ground structure is shown in Fig.~\ref{fig:groundrefs3}. The prescribed force densities are $q_b=5~\mathrm{kN/m}$, $q_i=1~\mathrm{kN/m}$, and $q_d=1~\mathrm{kN/m}$.}
    \label{fig:optres3}
\end{figure*}

The results for the three examples are shown in Figs.~\ref{fig:optres1}, \ref{fig:optres2}, and~\ref{fig:optres3}. For each example, results are presented for two reference cable network resolutions: an $11\times11$ grid and a $21\times21$ grid.
The optimized topologies exhibit pronounced holes where cable elements have been removed to satisfy the structural volume constraint. This member removal changes the force density distribution and, in turn, alters the optimized equilibrium form. Nevertheless, by construction, the solver seeks the cable network form that remains as close as possible to the reference configuration.
The popularity of form finding approaches for cable structures likely stems partly from their ability to generate nonintuitive structural layouts with a strong architectural and aesthetic impact. This characteristic is also evident in the approach proposed here: the optimized designs contain holes where elements have been removed, enhancing the architectural appeal of the resulting structures while reinforcing their lightweight character.

The optimized force density distributions are expected to converge to nearly discrete solutions in terms of the $\bm{\gamma}$ variables. These variables select the elements of the reference cable network that survive at the end of the optimization process. To promote convergence toward crisp final values of zero or one, two numerical strategies are adopted (see Sec.~\ref{sec:topopt}): SIMP penalization and the additional term $J_{\mathrm{bin}}$ in the objective function. Nevertheless, some variables do not converge exactly to zero or one. The $\bm{\gamma}$ variables obtained at the end of each topology optimization analysis are therefore rounded, and the corresponding equilibrated form is recomputed using the linear FDM.
After rounding the $\bm{\gamma}$ variables, the nodal deviations of both the optimized and rounded layouts are computed relative to the initial reference nodal positions, $\mathbf{r}_0$. The deviation between the optimized and rounded layouts is also computed to assess how much the cable network form obtained after rounding differs from that obtained directly through topology optimization.

\begin{table}[htbp]
    \centering
    \caption{Rounding study for the cable network of Example 1 with an $11\times11$ nodal grid (Fig.~\ref{fig:resex11}).}
    \label{tab:rounding-ex1_11x11_V040}
    \begin{tabular}{lccc}
        \toprule
        & Elements kept & $\frac{V_\mathrm{opt}}{V_0}$ & $\frac{V_\mathrm{round}}{V_0}$ \\
        \midrule
         & 168/420 (40.0\%) & 0.3586 & 0.3574 \\
        \bottomrule
    \end{tabular}
    \vfill
    \begin{tabular}{lccc}
        \toprule
        Nodal deviation & Mean & RMS & Maximum \\
         &  [m] &  [m] &  [m] \\
        \midrule
        Optimized vs.\ reference & 0.4227 & 0.4995 & 0.9495 \\
        Rounded vs.\ reference   & 0.4305 & 0.5032 & 0.9416 \\
        Rounded vs.\ optimized   & 0.0188 & 0.0349 & 0.1427 \\
        \bottomrule
    \end{tabular}
\end{table}

\begin{figure}[htbp]
    \centering
    \includegraphics[width=0.9\linewidth]{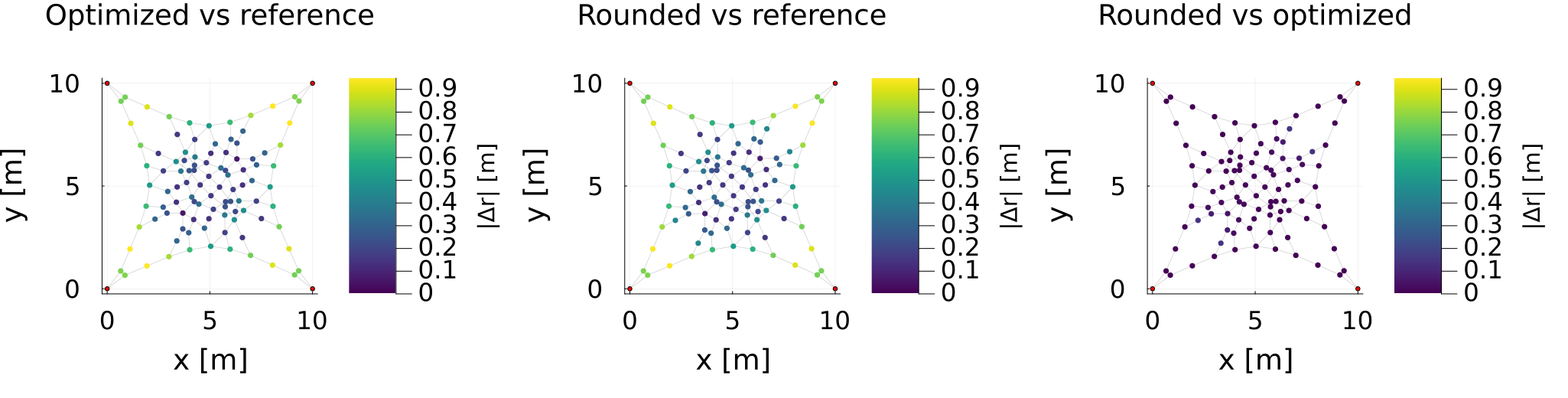}
    \caption{Nodal deviation contours for Example~1 with the $11\times11$ grid, where node color indicates the magnitude of deviation between (a) the optimized and reference configurations, (b) the rounded and reference configurations, and (c) the optimized and rounded configurations.}
    \label{fig:noderounded1}
\end{figure}

\begin{figure}[htbp]
    \centering
    \includegraphics[width=0.75\linewidth]{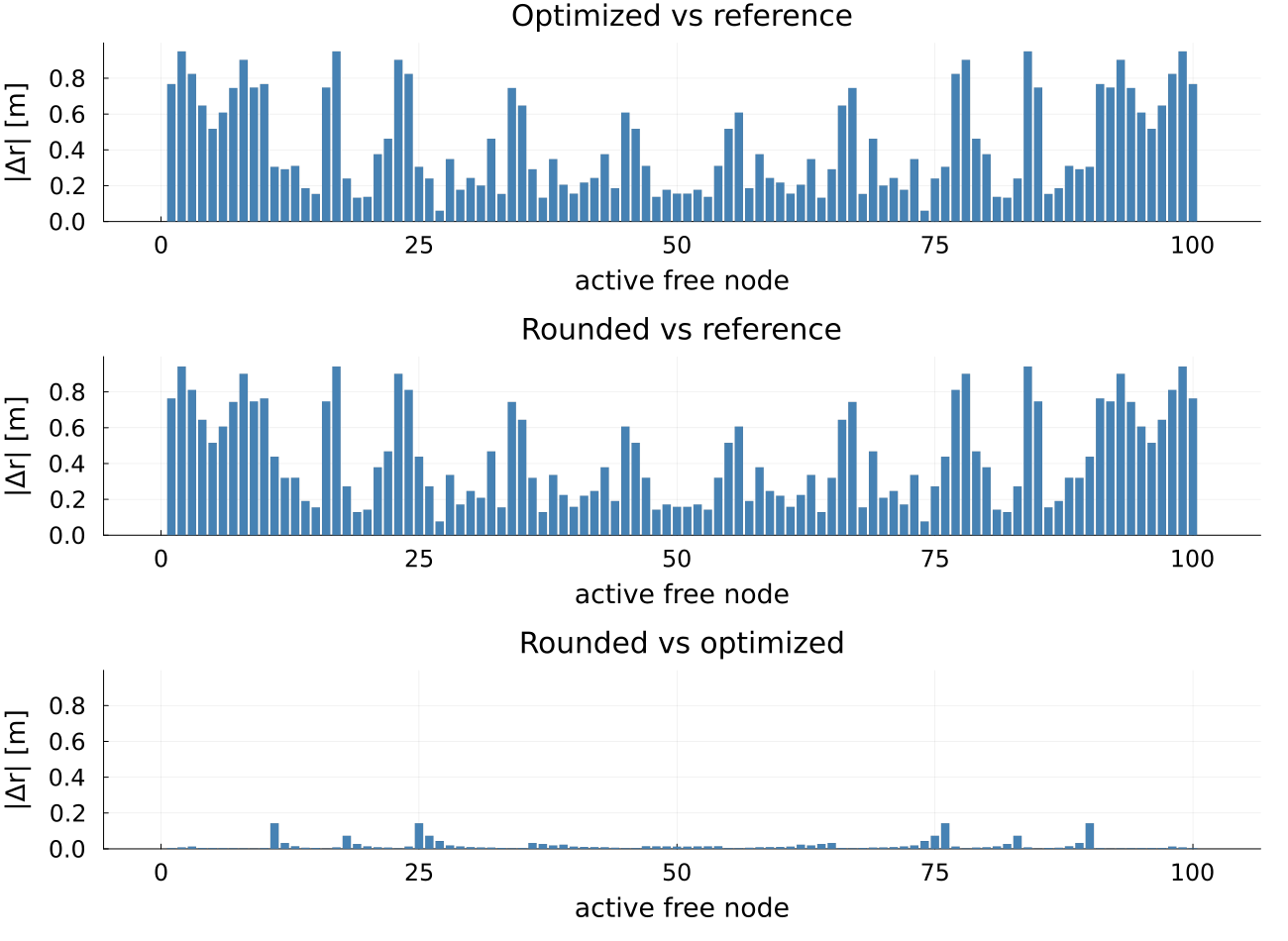}
    \caption{Bar plot of the nodal deviations for Example~1 with the $11\times11$ grid, comparing the optimized layout with the reference layout, the rounded layout with the reference layout, and the optimized layout with the rounded layout.}
    \label{fig:barrounded1}
\end{figure}

\begin{table}[htbp]
    \centering
    \caption{Rounding study for the cable network of Example 1 with a $21\times21$ nodal grid (Fig.~\ref{fig:resex12}).}
    \label{tab:rounding-ex1_21x21_V040}
    \begin{tabular}{lccc}
        \toprule
        & Elements kept & $\frac{V_\mathrm{opt}}{V_0}$ & $\frac{V_\mathrm{round}}{V_0}$ \\
        \midrule
         & 641/1640 (39.1\%) & 0.3231 & 0.3208 \\
        \bottomrule
    \end{tabular}
    \vfill
    \begin{tabular}{lccc}
        \toprule
        Nodal deviation & Mean & RMS & Maximum \\
         &  [m] &  [m] &  [m] \\
        \midrule
        Optimized vs.\ reference & 0.2874 & 0.3861 & 1.1720 \\
        Rounded vs.\ reference   & 0.2874 & 0.3861 & 1.1721 \\
        Rounded vs.\ optimized   & 0.0000 & 0.0000 & 0.0002 \\
        \bottomrule
    \end{tabular}
\end{table}

\begin{figure}[htbp]
    \centering
    \includegraphics[width=0.9\linewidth]{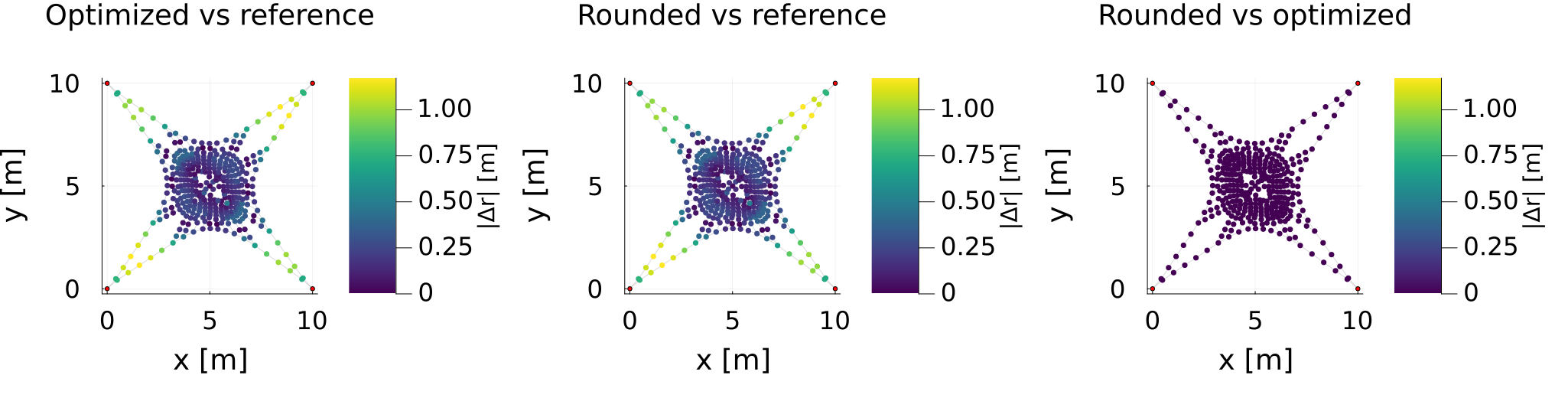}
    \caption{Nodal deviation contours for Example~1 with the $21\times21$ grid, where node color indicates the magnitude of deviation between (a) the optimized and reference configurations, (b) the rounded and reference configurations, and (c) the optimized and rounded configurations.}
    \label{fig:noderounded2}
\end{figure}

\begin{figure}[htbp]
    \centering
    \includegraphics[width=0.75\linewidth]{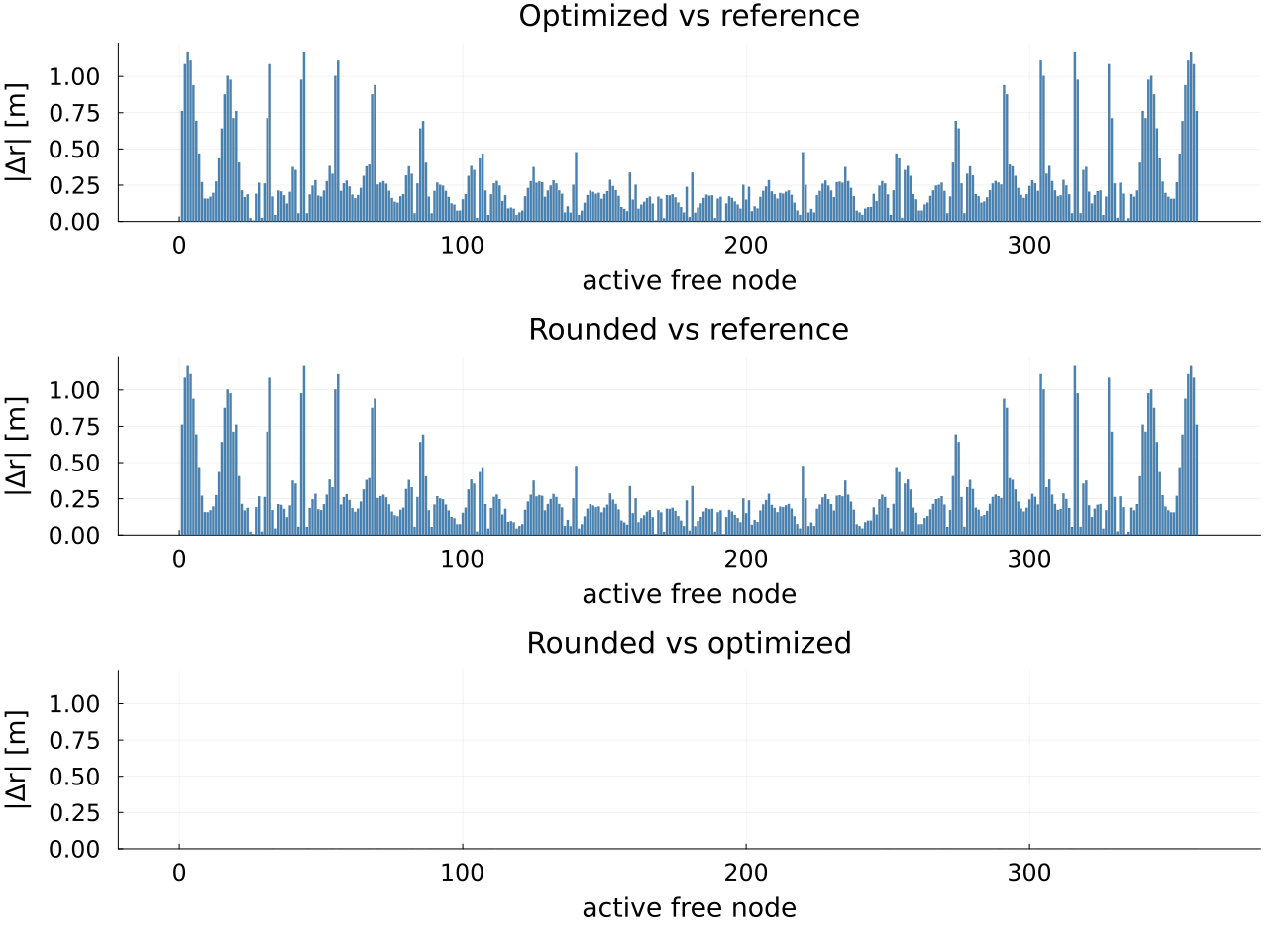}
    \caption{Bar plot of the nodal deviations for Example~1 with the $21\times21$ grid, comparing the optimized layout with the reference layout, the rounded layout with the reference layout, and the optimized layout with the rounded layout.}
    \label{fig:barrounded2}
\end{figure}

Tables~\ref{tab:rounding-ex1_11x11_V040} and~\ref{tab:rounding-ex1_21x21_V040} report the numerical results of the rounding study for Example~1 using the $11\times11$ and $21\times21$ reference grids, respectively. The tables present two complementary measures of design sparsity. The first, the fraction of elements kept, is a topological count. Of the $m$ members in the full ground structure, an element $e$ is retained if its optimized density exceeds the threshold $\gamma_e > 0.5$, and the reported percentage is $n_\mathrm{kept}/m$. This measure treats every member equally, regardless of its physical length. The second measure, relative volume, weights each member by its reference length $L_e$, which is equivalent to its material volume for a uniform cross section. Thus, $V_0 = \sum_e L_e$ for the full ground structure, $V_\mathrm{opt} = \sum_e \gamma_e L_e$ for the continuous optimized design, and $V_\mathrm{round} = \sum_{e \in \mathrm{kept}} L_e$ for the discrete topology obtained after thresholding. Because the ground structure contains members of different lengths, such as short orthogonal members and longer diagonal members, the percentage of elements retained and the volume fraction $V_\mathrm{round}/V_0$ may not coincide. A design can retain a large number of elements that collectively represent a small fraction of the total material volume, or vice versa. The close agreement between $V_\mathrm{opt}/V_0$ and $V_\mathrm{round}/V_0$ indicates that thresholding at $\gamma_e = 0.5$ closely preserves the volume of the continuous solution, even though it does not preserve the percentage of elements retained.

Tables~\ref{tab:rounding-ex1_11x11_V040} and~\ref{tab:rounding-ex1_21x21_V040} show that rounding the $\bm{\gamma}$ variables introduces only a minor perturbation to the optimized cable network form. For the $11\times11$ grid, the nodal deviation between the rounded and optimized configurations (mean $0.019$~m, maximum $0.143$~m) is nearly an order of magnitude smaller than the deviation introduced by the optimization itself, measured between the optimized and reference configurations (mean $0.423$~m, maximum $0.950$~m). For the $21\times21$ grid, the rounding step is essentially exact, with mean and RMS nodal deviations of $0.0$~m and a maximum deviation of only $0.0002$~m. This result indicates that the relaxed density variables converge to nearly discrete values without requiring post processing or correction. Interestingly, refinement of the reference ground structure appears to improve the discreteness of the optimized topology, at least in this example. At the same time, the fraction of elements kept and the corresponding volume fraction $V_\mathrm{round}/V_0$ do not coincide, and the gap between the two measures widens with mesh refinement. The difference increases from about $4$ percentage points at the $11\times11$ resolution ($40.0\%$ of elements kept compared with a $35.7$--$35.9\%$ volume fraction) to about $7$ percentage points at the $21\times21$ resolution ($39.1\%$ of elements kept compared with a $32.1$--$32.3\%$ volume fraction). This trend is consistent with the passive boundary elements, which are individually shorter at higher resolutions but are always retained. They therefore contribute a decreasing share of the total volume while still accounting for a fixed share of the element count. Finally, in both cases, the achieved volume fraction $V_\mathrm{opt}/V_0$ remains below the prescribed limit of $0.4$. Thus, the volume constraint is not active at the optimum, and the objective of matching the reference form can be satisfied without using the full material budget.

Figs.~\ref{fig:barrounded1} and~\ref{fig:barrounded2} show bar plots of the nodal deviations for the optimized and reference, rounded and reference, and rounded and optimized configurations. Figs.~\ref{fig:noderounded1} and~\ref{fig:noderounded2} show the corresponding contour plots for the same three comparisons. The highest deviations are concentrated near the perimeter, close to the fixed boundary nodes. The exception is the comparison between the rounded and optimized configurations for the $21\times21$ grid of Example~1, for which the deviation is negligible.

\subsection{Assessment of the volume fraction constraint}
\label{subsec:volcon}
This section assesses the effect of the volume constraint on the optimization results. The cable net of Example~2 (Fig.~\ref{fig:groundrefs2}), comprising $21\times21$ nodes, is considered. The allowable structural volume is varied by assigning different values to the upper bound $V^*$ in the constraint defined by Eq.~\eqref{eq:volconstr}.

\begin{figure}[htbp]
    \centering
    \includegraphics[width=0.9\linewidth]{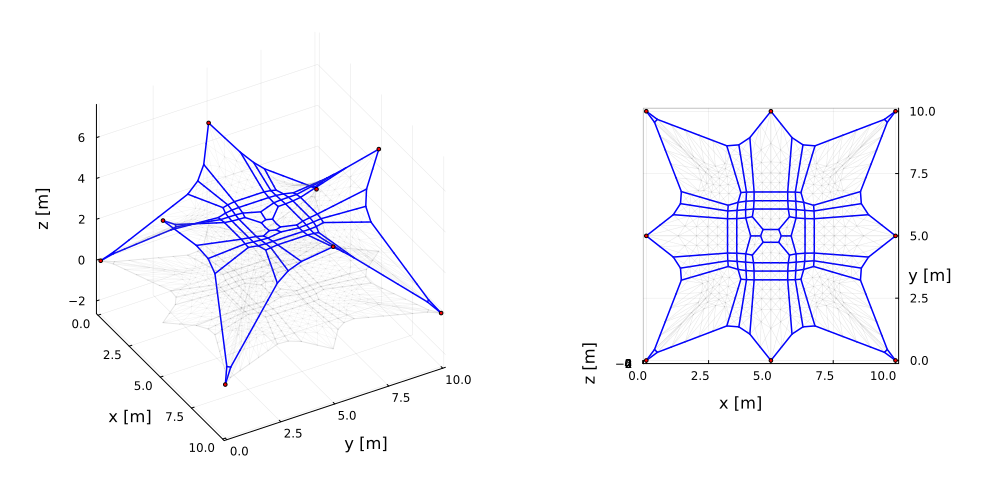}
    \caption{Force density topology optimization results for Example~2 with the $21\times21$ grid. The prescribed force densities are $q_b=3~\mathrm{kN/m}$, $q_i=2~\mathrm{kN/m}$, and $q_d=1~\mathrm{kN/m}$. The upper bound on the volume constraint is $V^*=20\%$.}
    \label{fig:topopt2V1}
\end{figure}

\begin{figure}[htbp]
    \centering
    \includegraphics[width=0.85\linewidth]{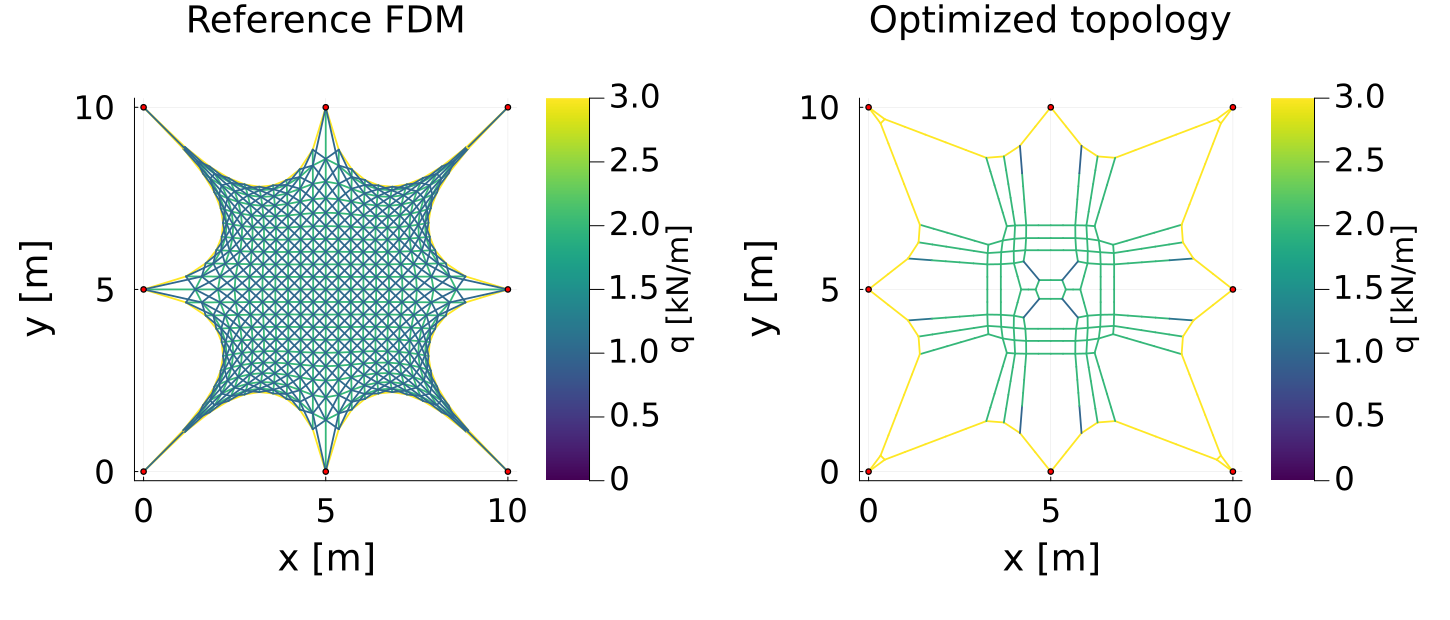}
    \caption{Force density in each element of Example~2 with the $21\times21$ grid, shown for both the reference structural form and the optimized configuration. The upper bound on the volume constraint is $V^*=20\%$.}
    \label{fig:qopt2V1}
\end{figure}

\begin{figure}[htbp]
    \centering
    \includegraphics[width=0.85\linewidth]{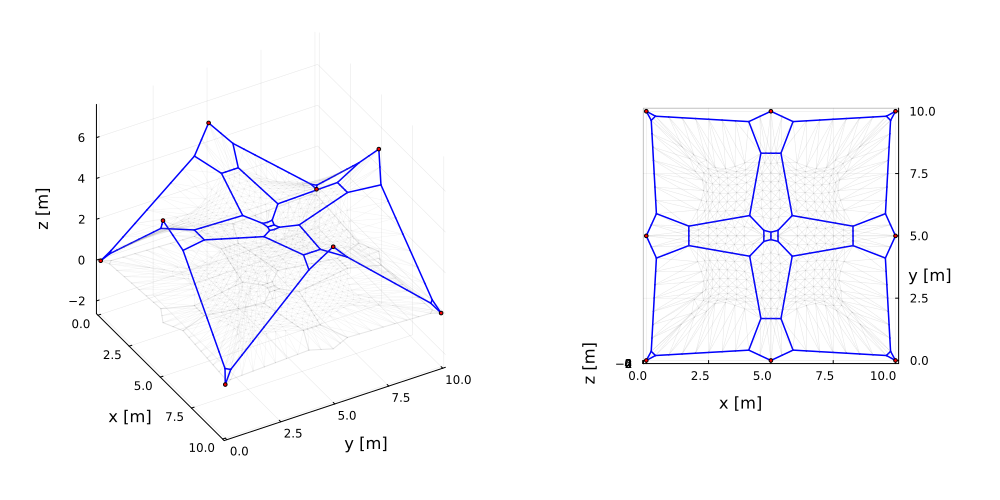}
    \caption{Force density topology optimization results for Example~2. The prescribed force densities are $q_b=3~\mathrm{kN/m}$, $q_i=2~\mathrm{kN/m}$, and $q_d=1~\mathrm{kN/m}$. The upper bound on the volume constraint is $V^*=15\%$.}
    \label{fig:topopt2V2}
\end{figure}

\begin{figure}[htbp]
    \centering
    \includegraphics[width=0.85\linewidth]{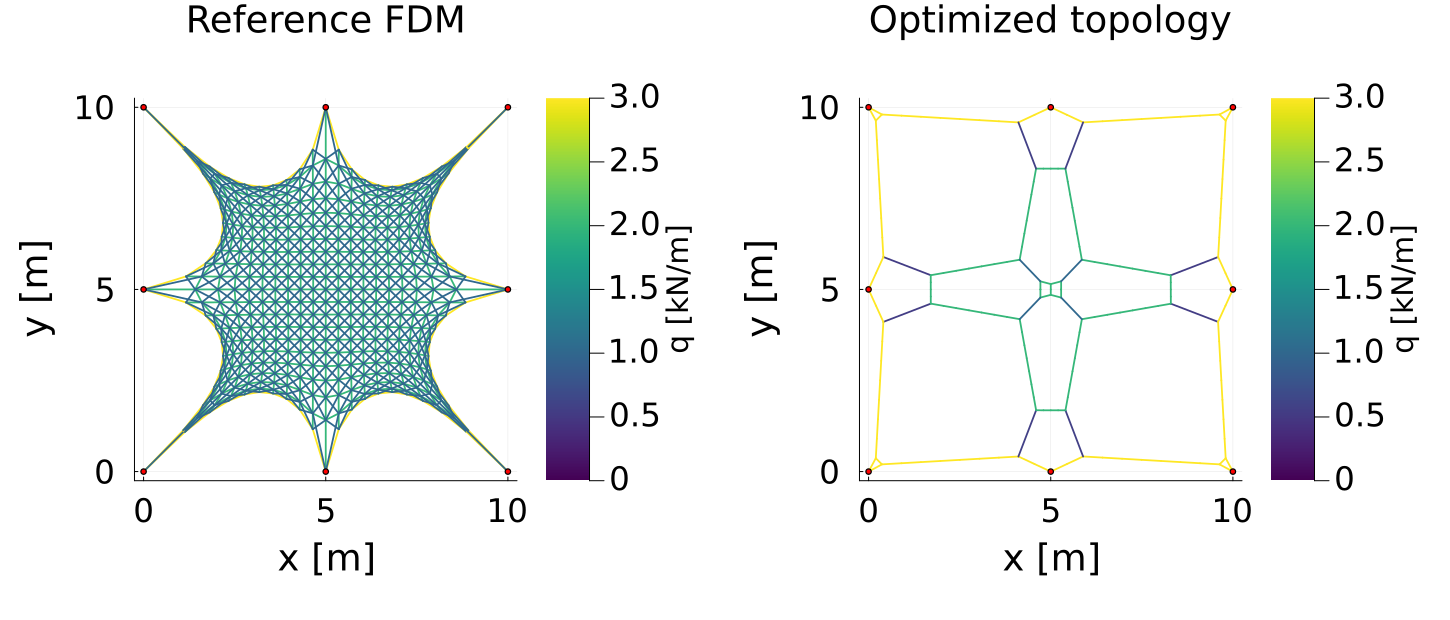}
    \caption{Force density in each element of Example~2 with the $21\times21$ grid, shown for both the reference structural form and the optimized configuration. The upper bound on the volume constraint is $V^*=15\%$.}
    \label{fig:qopt2V2}
\end{figure}

Let $V_0$ denote the volume of the corresponding reference cable structure. Reducing the allowable volume fraction from $V^*=0.40V_0$ to $V^*=0.20V_0$ and then to $V^*=0.15V_0$ produces a marked qualitative change in the optimized topology of Example~2, rather than merely reducing the number of elements in the reference cable net. At $V^*=0.20V_0$, the optimized configuration retains a redundant internal bracing pattern, with several intersecting orthogonal and diagonal members connecting the central region to the supports (Figs.~\ref{fig:topopt2V1} and~\ref{fig:qopt2V1}). At $V^*=0.15V_0$, this redundancy is essentially eliminated, and the optimized topology reduces to a minimal, nearly statically determinate load path. This path consists of a simple cross of direct members connecting the loaded joints to the supports (Figs.~\ref{fig:topopt2V2} and~\ref{fig:qopt2V2}). In both cases, the passive boundary elements are unaffected by the change in $V^*$, as expected from the constraint $\gamma_e=1$ for $e\in\mathcal{B}$, and continue to define the same outer envelope regardless of the interior volume budget.

\begin{figure}[htbp]
    \centering
    \includegraphics[width=0.9\linewidth]{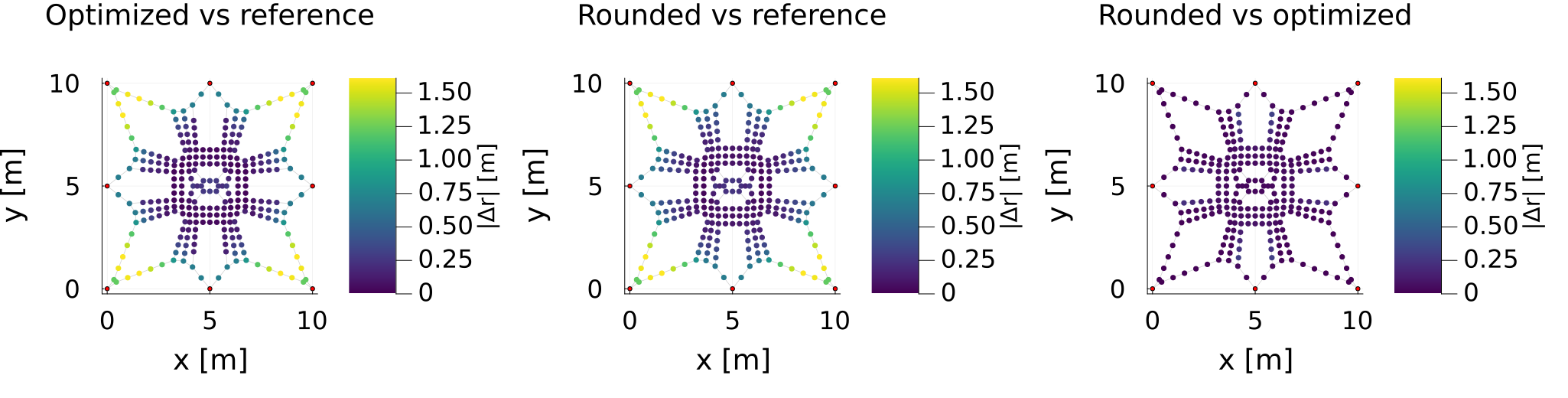}
    \caption{Nodal deviation contours for Example~2 with the $21\times21$ grid. The node color indicates the magnitude of the deviation between (a) the optimized and reference configurations, (b) the rounded and reference configurations, and (c) the optimized and rounded configurations. The upper bound on the volume constraint is $V^*=20\%$.}
    \label{fig:noderounded2V1}
\end{figure}

\begin{figure}[htbp]
    \centering
    \includegraphics[width=0.85\linewidth]{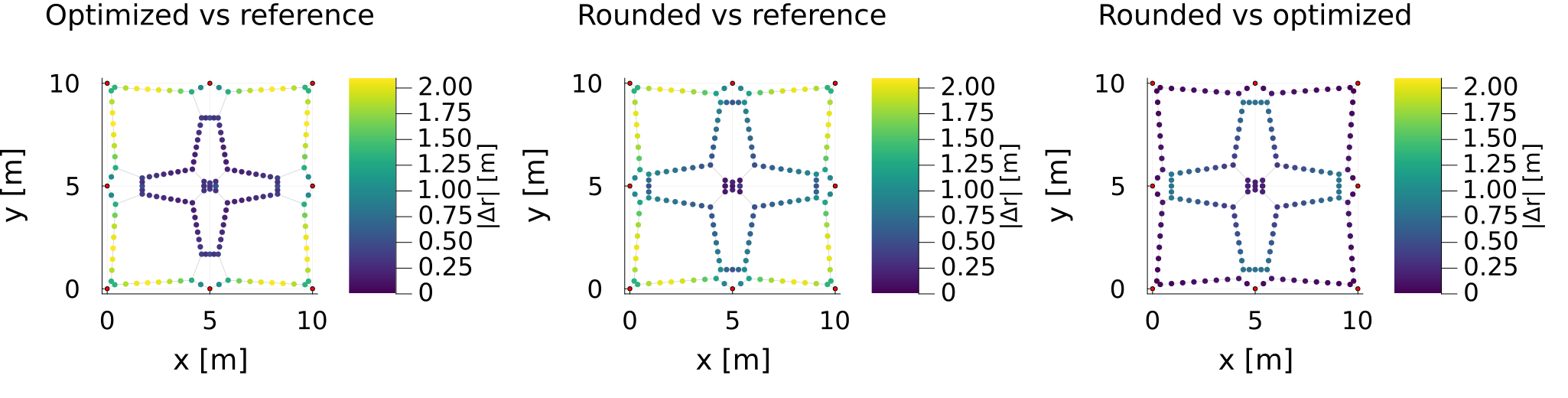}
    \caption{Nodal deviation contours for Example~2 with the $21\times21$ grid. The node color indicates the magnitude of the deviation between (a) the optimized and reference configurations, (b) the rounded and reference configurations, and (c) the optimized and rounded configurations. The upper bound on the volume constraint is $V^*=15\%$.}
    \label{fig:noderounded2V2}
\end{figure}

The nodal deviation contours in Figs.~\ref{fig:noderounded2V1} and~\ref{fig:noderounded2V2} are consistent with this change in topology. At $V^*=0.20V_0$, the largest deviations from the reference configuration are confined to the tips of the radiating diagonal members, while the densely connected central cluster remains close to $\mathbf{r}_0$. At $V^*=0.15V_0$, large deviations extend over most of the domain, and only a narrow band of nodes along the surviving cross shaped load path remains close to the reference form. This behavior reflects the decreasing number of active members available to reproduce the reference geometry as the volume budget is reduced. In both cases, the deviation between the rounded and optimized configurations remains markedly smaller than the deviation of either configuration from the reference. This result confirms that the nearly discrete convergence of $\bm{\gamma}$ observed for Example~1 also occurs for Example~2 under a substantially smaller volume fraction.

\subsection{Computational aspects of the proposed approach}
\label{subsec:comasptmdnlp}

This section investigates the computational performance of the interior point optimization solver MadNLP for a series of numerical examples. The analysis examines the relationship between problem size, expressed in terms of the numbers of variables and constraints, and the iterations and wall clock time required for convergence.

Table~\ref{tab:madnlp-summary} summarizes the problem size, number of iterations, and wall clock time required by MadNLP to solve Examples~1--3 using the $11\times11$, $21\times21$, and $31\times31$ meshes. As expected, the numbers of variables and constraints increase substantially with mesh refinement. For all three examples, they increase by approximately a factor of $8.5$ from the $11\times11$ grid to the $31\times31$ grid. The wall clock time increases considerably faster than the problem size, by approximately one to two orders of magnitude over the same refinement range. For example, it increases from $0.34$~s to $19.33$~s for Example~1 and from $0.65$~s to $61.64$~s for Example~3. This behavior reflects the increasing computational cost of each interior point iteration for larger systems and, in most cases, the greater number of iterations required at finer resolutions.

The number of iterations required for convergence does not increase monotonically with mesh size and varies considerably among the examples, ranging from $86$ for Example~2 with the $21\times21$ grid to $725$ for Example~2 with the $31\times31$ grid. This variability reflects perhaps the nonconvex nature of the underlying bilinear optimization problem. Its convergence behavior is sensitive to the particular combination of boundary conditions, force density assignments, and connectivity that defines each example, rather than to mesh resolution alone. Nevertheless, all cases converged within the prescribed iteration limit. Even the largest problem considered, with more than $17{,}000$ variables and $13{,}000$ constraints, was solved in approximately one minute on a standard desktop machine\footnote{All computations were performed on a MacBook Air with an Apple M4 chip and 32~GB of RAM.}. These results confirm the computational efficiency of the proposed formulation.
\begin{table}[htbp]
    \centering
    \caption{Performance of the MadNLP interior point solver: problem size, number of iterations, and wall clock time for the $11\times11$, $21\times21$, and $31\times31$ meshes of Examples~1--3.}
    \label{tab:madnlp-summary}
    \begin{tabular}{llrrrr}
        \toprule
        Example & Mesh & Variables & Constraints & Iterations & Wall time [s] \\
        \midrule
         Example 1 & $11\times11$ & 2031 & 1664 & 94 & 0.34 \\
         Example 1 & $21\times21$ & 7871 & 6324 & 339 & 5.69 \\
         Example 1 & $31\times31$ & 17511 & 13984 & 426 & 19.33 \\
         \hline
         Example 2 & $11\times11$ & 2019 & 1652 & 193 & 0.58 \\
         Example 2 & $21\times21$ & 7859 & 6312 & 86 & 1.39 \\
         Example 2 & $31\times31$ & 17499 & 13972 & 725 & 31.75 \\
         \hline
         Example 3 & $11\times11$ & 2028 & 1661 & 199 & 0.65 \\
         Example 3 & $21\times21$ & 7868 & 6321 & 335 & 5.58 \\
         Example 3 & $31\times31$ & 17508 & 13981 & 582 & 61.64 \\
        \bottomrule
    \end{tabular}
\end{table}

\subsection{Topology optimization results with additional form constraints}
\label{subsec:topoptWnonlindfdm}

This section presents numerical results for the topology optimization of a cable structure subject to additional nonlinear constraints that enforce prescribed geometric requirements. In particular, the proposed approach to topology optimized form finding is coupled with the nonlinear FDM. The nonlinear FDM described in Sec.~\ref{subsec:nonlinfdm} generates the reference form and, consequently, the ground structure used in the topology optimization.
This coupling affects both the initial form finding phase and the formulation of the topology optimization problem. The nonlinear geometric constraints imposed during the initial form finding analysis must also be retained during topology optimization.

The removal of cable elements during topology optimization changes the force densities and the corresponding equilibrium configuration. The optimized structure must therefore remain as close as possible to the reference form while satisfying the prescribed nonlinear geometric constraints. Consequently, the optimization problem considered here differs slightly from Problem~\eqref{eq:optprob2} and is formulated as follows:
\begin{equation}\label{eq:optprob3}
\begin{split}
\minimize_{\bm{\gamma},\mathbf{x},\mathbf{y},\mathbf{z}} \quad & J(\bm{\gamma},\mathbf{x},\mathbf{y},\mathbf{z})
= \left\|
\mathbf{r} - \mathbf{r}_0
\right\|_2^2 + \bm{\gamma}^T(\mathbf{1} - \bm{\gamma}) \\
\text{subject to } & \mathbf{q}(\bm{\gamma}) = \mathbf{q}_{min} +  \bm{\Gamma}^p \, (\mathbf{q}_0 - \mathbf{q}_{min})\\
        & \mathbf{C}^T_f \, \mathbf{Q}(\bm{\gamma}) \, \left(\mathbf{C}_f \, \mathbf{x}_f +  \mathbf{C}_c \, \mathbf{x}_c \right) = \mathbf{f}_x\\
        & \mathbf{C}^T_f \, \mathbf{Q}(\bm{\gamma}) \, \left(\mathbf{C}_f \, \mathbf{y}_f +  \mathbf{C}_c \, \mathbf{y}_c \right) = \mathbf{f}_y\\
        & \mathbf{C}^T_f \, \mathbf{Q}(\bm{\gamma}) \, \left(\mathbf{C}_f \, \mathbf{z}_f +  \mathbf{C}_c \, \mathbf{z}_c \right) = \mathbf{f}_z.\\
        & \sum_{i=1}^m \gamma_i l_{0,i} \leq V^* \\
        & \sum_{i \in \mathcal{E}_a} \gamma_i^p \geq \gamma_L^{\min}, \; \forall \; a \in \mathcal{N}_L\\
        & \left\|\mathbf{r}_a-\mathbf{r}_b\right\|_2^2=l_{0,e}^2,
      \quad e=(a,b)\in\mathcal{E}_c,\\
        & \gamma_i \in [0,1] \text{ for } i=1,\dots,m \\
        & \gamma_i = 1 \; \forall \; i \in \mathcal{B}\\
        & \gamma_i = 1 \; \forall \; i =(a,b)\in \mathcal{E}_c,
\end{split}
\end{equation}
Here, $\mathbf{r}_a=[x_a,y_a,z_a]^T$ is the position of node $a$, $\mathcal{E}_c$ is the set of members with prescribed lengths, and $l_{0,e}$ is the target length of member $e$. In Problem~\eqref{eq:optprob3}, the additional constraints preserve the original lengths of the members in $\mathcal{E}_c$ and require their design variables to remain equal to one. These members are therefore treated as passive and cannot be removed during topology optimization. The resulting nonlinear optimization problem is modeled using JuMP.jl and solved with Ipopt.jl.

\begin{figure}[htbp]
    \centering
    \includegraphics[width=0.9\textwidth]{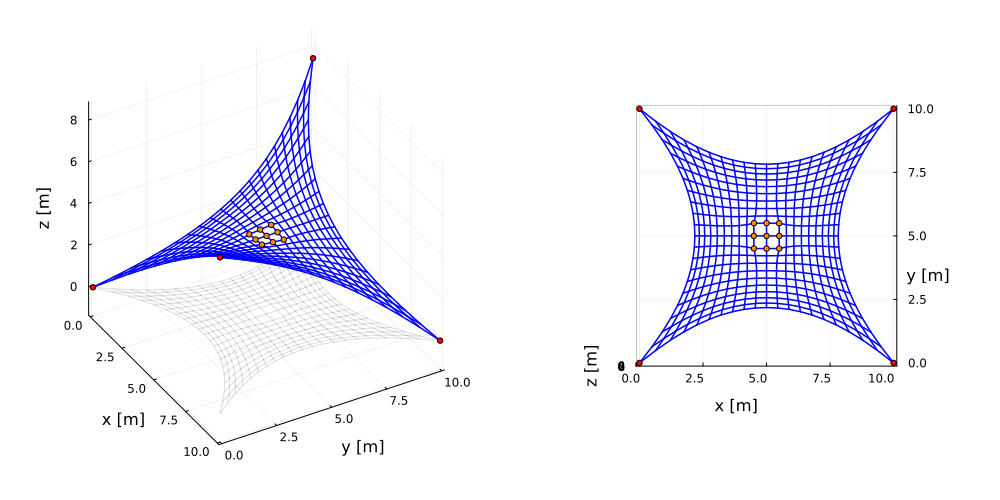}
    \caption{Reference form generated using the nonlinear FDM and used in Sec.~\ref{subsec:topoptWnonlindfdm}. The orange nodes identify the region in which the lengths of the members connecting pairs of marked nodes are constrained to remain equal to their original values, both during the nonlinear form finding analysis and the subsequent topology optimization.}
    \label{fig:refnonlinfdm04}
\end{figure}

\begin{figure}[htbp]
    \centering
    \includegraphics[width=0.9\textwidth]{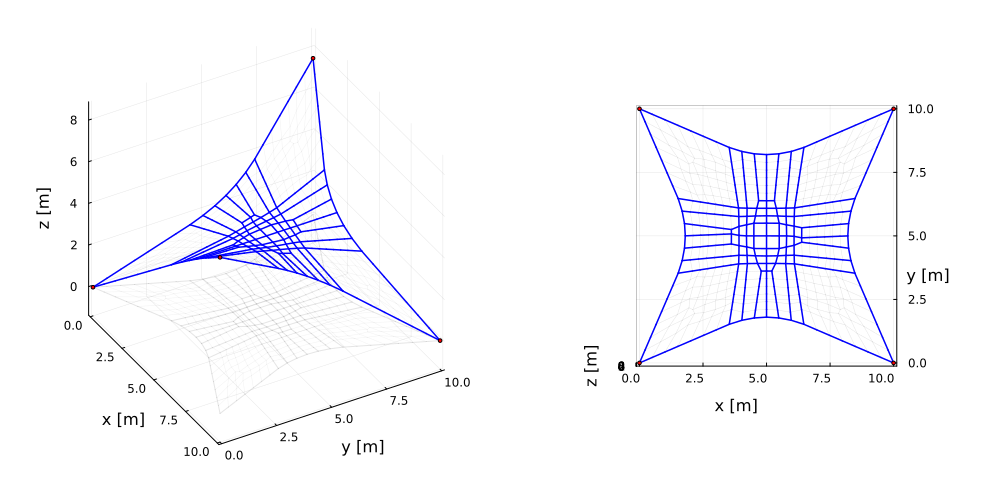}
    \caption{Topology optimization result obtained with the nonlinear form constraints and a prescribed volume fraction of $0.40$ in Sec.~\ref{subsec:topoptWnonlindfdm}.}
    \label{fig:topnonlinfdm040}
\end{figure}

\begin{figure}[htbp]
    \centering
    \includegraphics[width=0.9\textwidth]{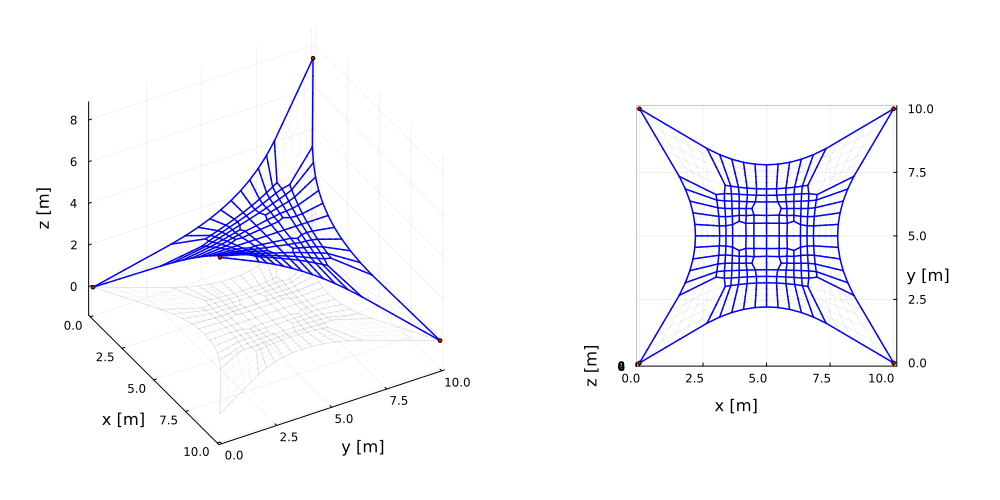}
    \caption{Topology optimization result obtained with the nonlinear form constraints and a prescribed volume fraction of $0.60$ in Sec.~\ref{subsec:topoptWnonlindfdm}.}
    \label{fig:topnonlinfdm060}
\end{figure}

Fig.~\ref{fig:refnonlinfdm04} shows the reference equilibrium form used as the ground structure for topology optimization. The initial equilibrium geometry supplied to the nonlinear FDM (see Problem~\eqref{eq:nlfdm-nlp}) is generated using the linear FDM with boundary and internal force densities of $q_b=5$ and $q_i=1$, respectively, as in Example~1. This calculation defines the initial force density vector $\mathbf{q}_0$ and the associated nodal coordinate vectors $\mathbf{x}_0$, $\mathbf{y}_0$, and $\mathbf{z}_0$. The orange nodes identify the region in which the members connecting pairs of marked nodes are constrained to retain their original lengths during both nonlinear form finding and the subsequent topology optimization.
The reference cable structure contains only members oriented along the horizontal and vertical directions in the projected $(x,y)$ plane; no diagonal members are included in the reference grid. This choice is based on preliminary numerical experiments in which the inclusion of diagonal members, together with the prescribed length constraints for the members connecting the orange nodes, did not yield feasible converged solutions. It was not possible to determine whether this behavior resulted from numerical difficulties encountered by Ipopt or from the infeasibility of the optimization problem. One possible explanation is that the addition of diagonal members increased the number of passive members whose design variables were fixed at $\gamma_i=1$. The resulting reduction in the design space may have prevented the optimizer from identifying a feasible solution.

Figs.~\ref{fig:topnonlinfdm040} and~\ref{fig:topnonlinfdm060} show the structural forms associated with the optimized force density distributions for allowable structural volume fractions of $0.40$ and $0.60$, respectively. In both cases, the optimizer identifies cable topologies whose equilibrium forms remain similar to the reference configuration in Fig.~\ref{fig:refnonlinfdm04}. Moreover, the members connecting pairs of orange nodes retain their prescribed original lengths, confirming that the additional geometric constraints are satisfied throughout the topology optimization process.

These results demonstrate that the proposed approach to form finding through force density topology optimization can accommodate additional constraints on the geometry, shape, and equilibrium configuration. Other requirements may therefore be incorporated, provided that they can be expressed in terms of the nodal coordinates or force densities, which are optimization variables in the proposed formulation.


\section{Conclusions}
\label{sec:end}

This paper presents a novel topology optimization approach to the form finding of cable structures based on the Force Density Method (FDM). The proposed formulation couples the linear equilibrium structure of the FDM with a density based topology optimization scheme.
Binary design variables $\bm{\gamma}$ associated with each cable element are relaxed into continuous densities and penalized through a SIMP interpolation of the force densities.
The approach is complemented by an explicit binary promoting term in the objective function. The resulting continuous optimization problem is solved with the interior point solver MadNLP.jl, using JuMP.jl to formulate the model and automatically differentiate the objective and constraint functions.

An optimization-based formulation of the nonlinear FDM is also proposed. In this formulation, the force densities and nodal coordinates are treated as optimization variables, while equilibrium and prescribed geometric requirements are imposed as nonlinear equality constraints. A force density vector and its associated equilibrium coordinates, obtained using the linear FDM, provide the initial point for the nonlinear optimization analysis. The objective minimizes the change in the force densities relative to this initial configuration. The nonlinear FDM problem is modeled using JuMP.jl and solved with Ipopt.jl using automatic differentiation.

The numerical examples showed that the proposed formulation could identify sparse cable net topologies that remained close to a prescribed reference configuration while satisfying a prescribed volume budget and maintaining minimum connectivity at the loaded joints. The optimized designs exhibited pronounced holes where cable elements were removed. These results illustrated how the redistribution of force densities following member removal reshaped the equilibrium geometry while retaining the general architectural character of the reference form.
A dedicated rounding study showed that the relaxed density variables converged to nearly discrete values at the end of the optimization process, with only a minor correction required after thresholding. The nodal deviation introduced by rounding was nearly an order of magnitude smaller than the deviation introduced by the optimization itself, and one case converged to an essentially exact discrete solution. Interestingly, the degree of discreteness improved with mesh refinement, at least in the numerical examples considered in this paper. The fraction of elements retained and the corresponding volume fraction did not coincide because of the different member lengths in the ground structure, and this difference increased with mesh refinement.

The assessment of the volume fraction constraint further showed that reducing the allowable structural volume did not simply reduce the reference cable net layout proportionally. Instead, it could trigger a qualitative transition in the optimized topology from a redundant, densely braced configuration to a minimal load path. This behavior highlighted the ability of the proposed formulation to capture nontrivial and nonintuitive structural layouts, which are often sought in the design of architecturally expressive cable and tensile structures.
Finally, the study of the computational performance of MadNLP showed that the proposed formulation scaled reasonably well with problem size. The largest problems considered, comprising more than $17{,}000$ variables and $13{,}000$ constraints, were solved in approximately one minute on a standard desktop machine. The number of iterations required for convergence appeared to depend more strongly on the particular combination of boundary conditions, force densities, and connectivity in each example than on mesh resolution alone, possibly because of the nonconvex nature of the underlying optimization problem.

The topology optimization formulation was further applied to a reference configuration generated using the nonlinear FDM. The geometric constraints imposed during the nonlinear form finding analysis were retained during the topology optimization phase, and the members associated with these constraints were treated as passive. The resulting designs for structural volume fractions of $0.40$ and $0.60$ remained similar to the nonlinear reference form while preserving the prescribed member lengths. These results showed that the proposed framework could incorporate additional requirements affecting the geometry, force densities, and equilibrium configuration throughout both form finding and topology optimization.

Overall, the results demonstrated that the proposed force density topology optimization framework provided a simple, flexible, and computationally efficient tool for the form finding and topology optimization of cable net structures, including problems subject to additional geometric constraints. Future work may extend the present formulation to accommodate further engineering requirements within the form finding process, such as those considered by \citet{quagliaroli2013flexible}. The approach could also be extended to membrane and shell structural systems, providing a further step toward the topology optimized form finding of architectural minimal surface structures.

\section*{Replication of results}
\noindent The numerical and algorithmic details are provided
in sufficient detail in the text to be implemented. 

\section*{Conflict of interest}
\noindent The author declares that he has no conflict of interest.

\section*{Acknowledgments}
\noindent NP acknowledges the support of the Israel Science Foundation (ISF), grant No. 304/24.


\bibliography{mybibfile}

\end{document}